\pdfoutput=1
\documentclass[final]{siamonline0516}
\usepackage{subfigmat}% matrices of similar subfigures, aka small multiples
\usepackage{amsmath,amsfonts,mathtools}
\usepackage{color}
\usepackage{soul}
\usepackage[algo2e, linesnumbered]{algorithm2e}
\usepackage{bm}

\graphicspath{{./}{./figs/}}

  \def\clap#1{\hbox to 0pt{\hss#1\hss}}

\providecommand{\mat}[1]{\bm{#1}}%
\renewcommand{\vec}[1]{\mathbf{#1}}

\providecommand{\mA}{\ensuremath{\mat{A}}}

\providecommand{\mC}{\ensuremath{\mat{C}}}

\providecommand{\mG}{\ensuremath{\mat{G}}}

\providecommand{\mI}{\ensuremath{\mat{I}}}

\providecommand{\mK}{\ensuremath{\mat{K}}}

\providecommand{\mM}{\ensuremath{\mat{M}}}
\providecommand{\mN}{\ensuremath{\mat{N}}}

\providecommand{\mQ}{\ensuremath{\mat{Q}}}
\providecommand{\mR}{\ensuremath{\mat{R}}}

\providecommand{\mU}{\ensuremath{\mat{U}}}

\providecommand{\vf}{\ensuremath{\vec{f}}}
\providecommand{\vg}{\ensuremath{\vec{g}}}
\providecommand{\vh}{\ensuremath{\vec{h}}}

\providecommand{\vk}{\ensuremath{\vec{k}}}

\providecommand{\vm}{\ensuremath{\vec{m}}}

\providecommand{\vu}{\ensuremath{\vec{u}}}
\providecommand{\vv}{\ensuremath{\vec{v}}}

\providecommand{\vx}{\ensuremath{\vec{x}}}

\usepackage{amssymb}
\usepackage{bbm}
\usepackage{marginnote}
\newcommand\rsout{\bgroup\markoverwith
{\textcolor{red!70!black}{\rule[0.5ex]{2pt}{1.5pt}}}\ULon}

\title{Dimension Reduction via Gaussian Ridge Functions
} 
\author{Pranay Seshadri\thanks{Postdoctoral Fellow, Department of Engineering, University of Cambridge, U.K., and Group Leader, Data-Centric Engineering, The Alan Turing Institute, London, U. K., \texttt{ps583@cam.ac.uk, pseshadri@turing.ac.uk}} \and Shaowu Yuchi\thanks{Undergraduate Research Assistant, Department of Engineering, University of Cambridge, U. K.} \and Geoffrey T. Parks\thanks{Reader, Department of Engineering, University of Cambridge, U. K.} }

\begin{document}
\maketitle
\newcommand{\slugmaster}{%
\slugger{juq}{2016}{xx}{x}{x--x}}%slugger should be set to juq, siads, sifin, or siims

\begin{abstract}
Ridge functions have recently emerged as a powerful set of ideas for subspace-based dimension reduction. In this paper we begin by drawing parallels between ridge subspaces, sufficient dimension reduction and active subspaces, contrasting between techniques rooted in statistical regression and those rooted in approximation theory. This sets the stage for our new algorithm that approximates what we call a Gaussian ridge function---the posterior mean of a Gaussian process on a dimension-reducing subspace---suitable for both regression and approximation problems. To compute this subspace we develop an iterative algorithm that optimizes over the Stiefel manifold to compute the subspace, followed by an optimization of the hyperparameters of the Gaussian process. We demonstrate the utility of the algorithm on two analytical functions, where we obtain near exact ridge recovery, and a turbomachinery case study, where we compare the efficacy of our approach with three well-known sufficient dimension reduction methods: SIR, SAVE, CR. The comparisons motivate the use of the posterior variance as a heuristic for identifying the suitability of a dimension-reducing subspace. 
\end{abstract}



\pagestyle{myheadings}
\thispagestyle{plain}
\markboth{SESHADRI, YUCHI, PARKS}{Gaussian Ridge Functions}

\section{Introduction}
Dimension reduction is essential for modern data analysis. Its utility spans both simulation informatics and statistical regression fields. The objective of dimension reduction is to obtain parsimoniously parameterized models---functions of few variables that can be used for predicting, understanding and visualizing the trends observed in high-dimensional data sets. These data sets may originate from a tailored \emph{design of experiment} of computer models (see \cite{pukelsheim1993optimal}) and are therefore deterministic, or they may originate from the observations of sensors and are therefore subject to measurement noise. Across both paradigms, techniques and ideas within \emph{subspace-based dimension reduction} have proven to be extremely fruitful in inference. Application areas that have benefitted from subspace-based dimension reduction range from airfoil aerodynamics \cite{grey2017active} and combustion \cite{magri2016stability} to economic forecasting (see page 4403 in~\cite{Adragni2009}), turbomachinery aerothermodynamics \cite{seshadri2017turbo}, solar energy \cite{constantine2015discovering} and epidemiology \cite{Diaz2018141}. 

\begin{figure}
\begin{subfigmatrix}{3}% number of columns
\subfigure[]{\includegraphics{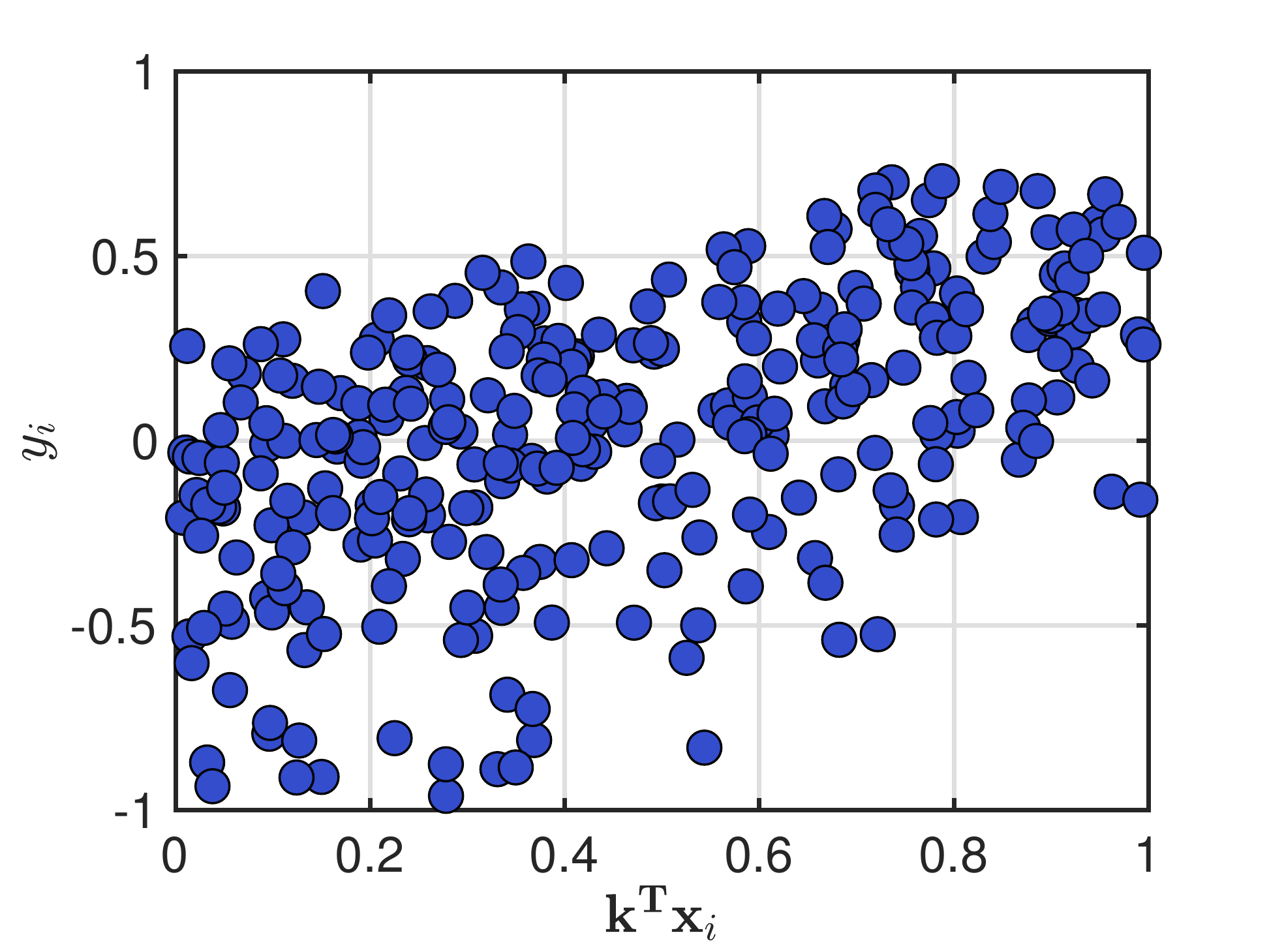}}
\subfigure[]{\includegraphics{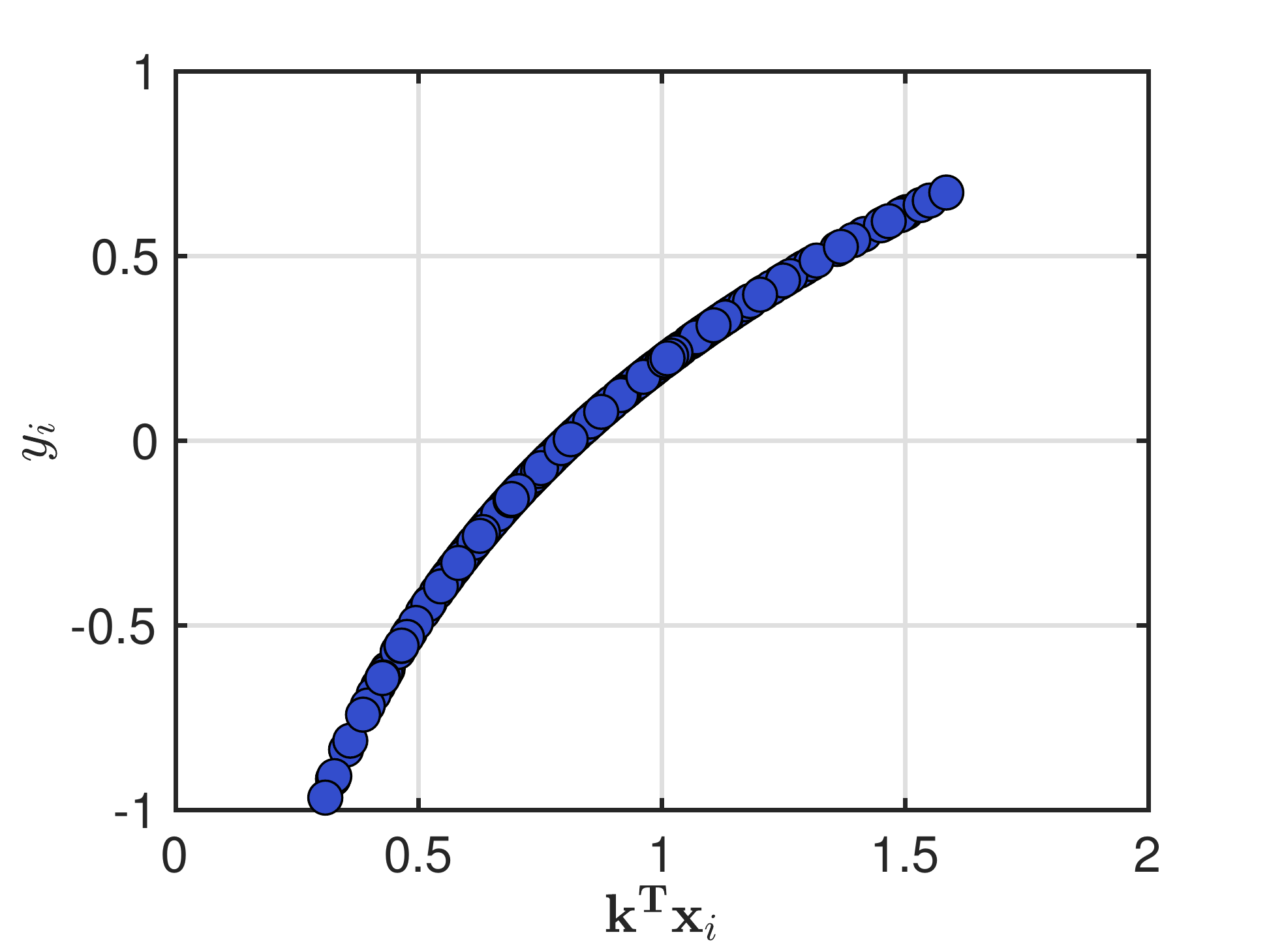}}
\subfigure[]{\includegraphics{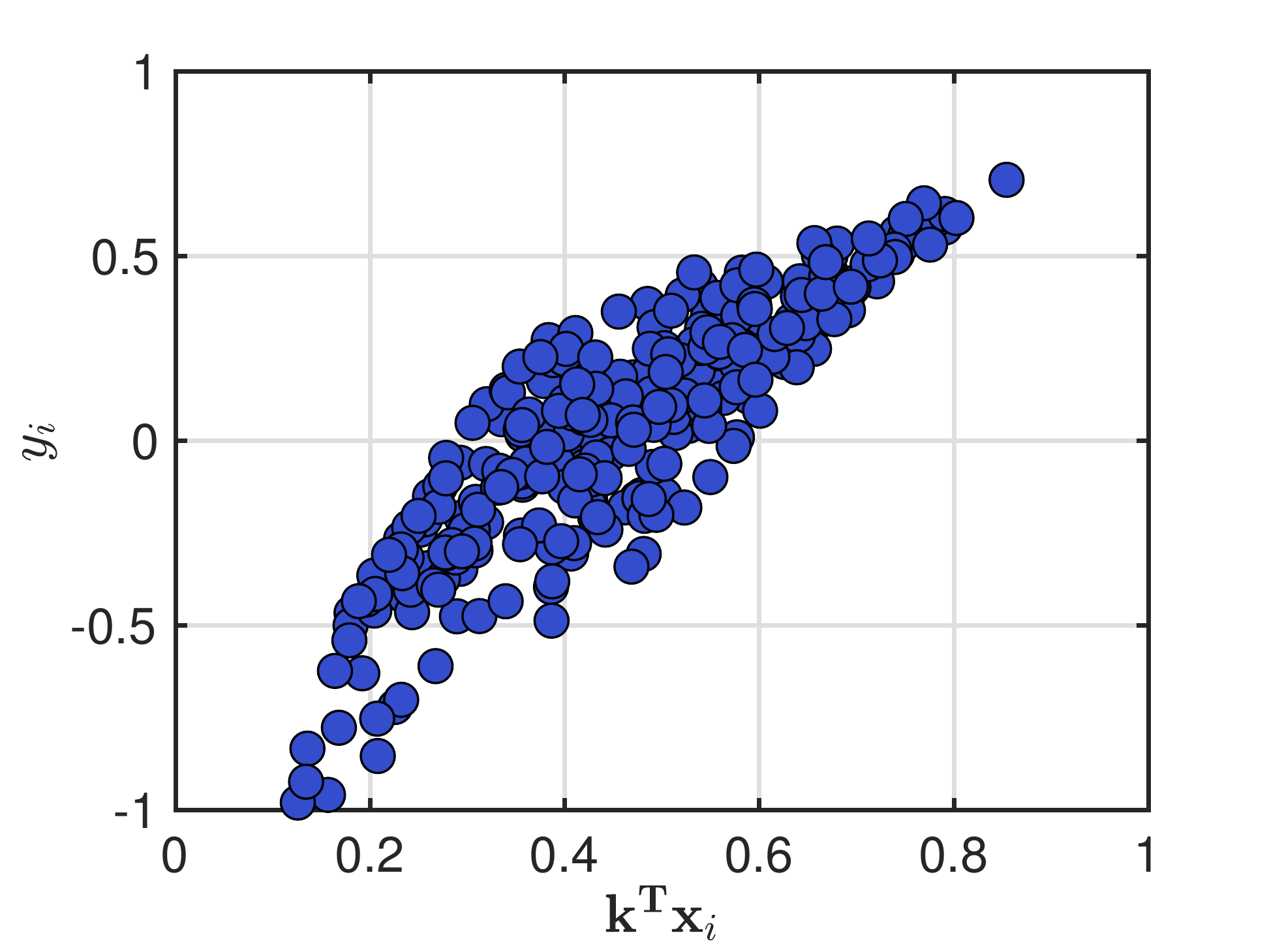}}
\end{subfigmatrix}
\caption{Sufficient summary plots of the function $y=log(x_1 + x_2 + x_3)$ where the horizontal axis is given by $\mathbf{k}^T (x_1, \; x_2, \; x_3)$. In (a) $\mathbf{k}=(1, 0, 0)^T$, in (b) $\mathbf{k}=(1/\sqrt{3}, 1/\sqrt{3}, 1/\sqrt{3})^T$, while in (c) $\mathbf{k}=(0.2, 0.5, 0.2)^T$.}
\label{figure_intro}
\end{figure}

To motivate an exposition of the relevant literature and to set the stage for our paper, we introduce the trivariate function $y = \text{log}(x_1 + x_2 + x_3)$.  In Figure~\ref{figure_intro}, we generate scatter plots of the function on three different subspaces $\mathbf{k}^T \vx$---i.e., linear combinations of the three variables $x_1$, $x_2$ and $x_3$. In (a) $\mathbf{k}$ is one of canonical vectors, in (b) all the elements of $\mathbf{k}$ are equivalent (and normalized), while in (c) we moderately perturb the entries in (b). Naturally, we find the subspace $\mathbf{k}^T \vx$ in (b) to be \emph{optimal} in the sense that it most accurately characterizes the 3D function as a function of one variable. Scatter plots of the kind in Figure~\ref{figure_intro}, where the horizontal axis is a subspace of the inputs, are known as \emph{sufficient summary plots} \cite{cook2009regression}. These plots are useful in identifying low-dimensional structure in high-dimensional problems. However, as the selection of $\mathbf{k}$ in Figure~\ref{figure_intro} illustrates, the impediment to realizing this identification is an appropriate choice of the subspace.

Enter \emph{active subspaces} \cite{constantine2015active}, a collection of ideas that facilitate subspace-based dimension reduction for deterministic computer models. If one can compute gradients of a function $f$, or even approximate them, then one can identify directions along which (on average) $f$ exhibits the greatest variation, and conversely directions along which (on average) $f$ is near constant\footnote{Assuming, of course, $f$ admits an active subspace; there is no guarantee that every differentiable multivariate function $f$ admits an active subspace.}. Active subspaces owes their development to initial work by Samarov \cite{samarov1993exploring} and more recent work by Constantine et al.~(see \cite{constantine2014computing, constantine2014active, constantine2015active}). The latter references build the theory of active subspaces and also offer practical algorithms for their computation. Parallels between active subspaces and \emph{ridge functions} have also been recently explored in the literature \cite{constantine2016anear}. 

The topic of ridge functions \cite{pinkus2015ridge} may be interpreted as a broad generalization of ideas within active subspaces. They are of the form $f(\vx) = g(\mM^T \vx)$, where $\mM$ is a tall matrix, implying that $g$ is a function of fewer variables than $f$. Functions of this form are known as \emph{generalized ridge functions}. When $d=1$, Pinkus \cite{pinkus2015ridge} defines these functions as \emph{ridge functions}. The elements of $\mM$ and the function $g$ are useful in identifying low-dimensional structures, in a similar vein to the example discussed above.  In the case where $\mM$ is a vector, which implies that $g$ is univariate function, Cohen et al.~\cite{cohen2012capturing} provide estimates on how well $f$ can be approximated using only point evaluations. Building on their work, Fornasier et al.~\cite{fornasier2012learning} study the case where $\mM$ is a matrix and provide two randomized algorithms for approximating $g(\mM^T \vx)$. The work of Tyagi and Cevher \cite{tyagi2014learning} is also similar in scope; they provide an algorithm that approximates functions of the form $f(\vx) = \sum_{i=1}^{n} g_i(\mM_{i}^T \vx)$. Clearly, there are similarities between ridge functions, projection pursuit regression (see page 390 in \cite{friedman2001elements}) and, by association, single layer neural networks. The projection pursuit regression problem can be interpreted as an approximation with
\begin{equation}
\sum_{i=1}^{k}g_{i}\left(\vm_{i}^{T} \vx  \right),
\end{equation}
where $\vm_{i} \in \mathbb{R}^{d}$ and the univariate function $g_{i}$ are unknown; determined via a stepwise greedy algorithm following Friedman and Stuetzle \cite{friedman1981projection}. It is worth emphasizing that although our function may not admit an analytical ridge structure, i.e., $f(\vx) \neq g(\mM^T \vx)$, one may still be able to approximate $f(\vx) \approx g(\mM^T \vx)$ extracting useful inference. We refer to this as a \emph{ridge approximation}, that is obtained by evaluations of $f$ and by minimizing a suitably constructed objective function to ascertain $\mM$ and $g$. 

In introducing methods for approximation and statistical regression (projection pursuit regression), it is important to note the differences between both paradigms. In statistical regression, as alluded to previously, $f(\vx)$ is not deterministic and therefore has a conditional density that is not a delta function \cite{constantine2015active}. Furthermore, the joint density of $\vx$ is not known in the regression setting. In comparison, when working with computer models---i.e., in the approximation paradigm---the joint density of $\vx$ is known and often prescribed according to some optimal design of experiment, for example, Latin hypercube sampling, D-optimal design, etc. Lastly, techniques for estimating gradients for unknown regression functions are notoriously expensive in high dimensions. Thus, non-gradient-based approaches are sought. 

The field of sufficient dimension reduction is rich in ideas for achieving subspace-based dimension reduction within a statistical regression framework. The goal here is to replace the vector of covariates (inputs) with their projection onto a subspace assembled from the space of the covariates themselves \cite{cook2005sufficient}. This subspace must be constructed without loss of information based either on the conditional mean, conditional variance or, more generally, the conditional distribution of the outputs with respect to the inputs. The numerous methods for estimating these subspaces include sliced inverse regression (SIR) \cite{li1991sliced}, sliced average variance estimation (SAVE) \cite{dennis2000save} and contour regression (CR) \cite{li2007directional}, to name but a few (see \cite{ma2013review} for a detailed review). 

\begin{figure}
\begin{subfigmatrix}{3}% number of columns
\subfigure[]{\includegraphics{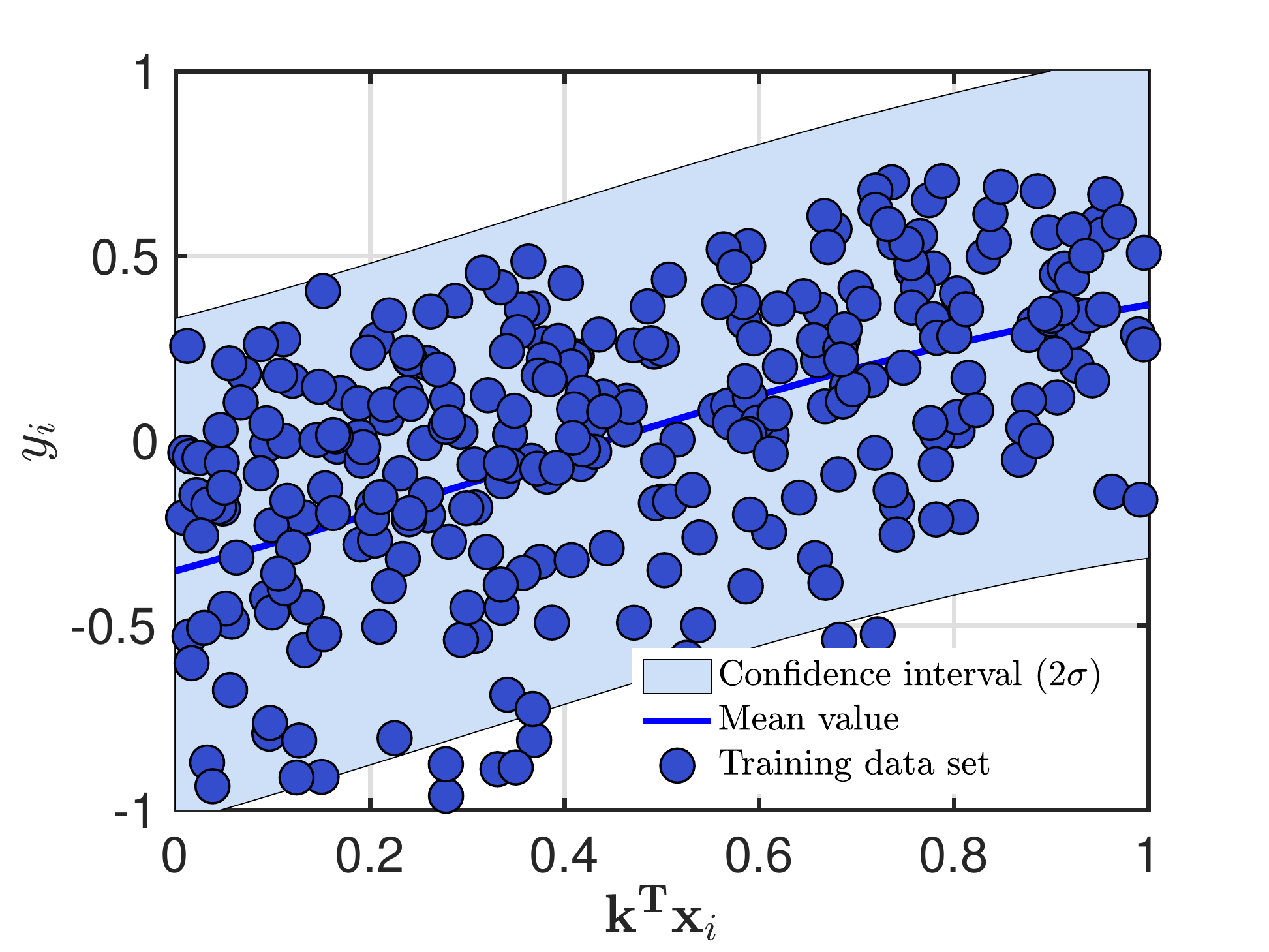}}
\subfigure[]{\includegraphics{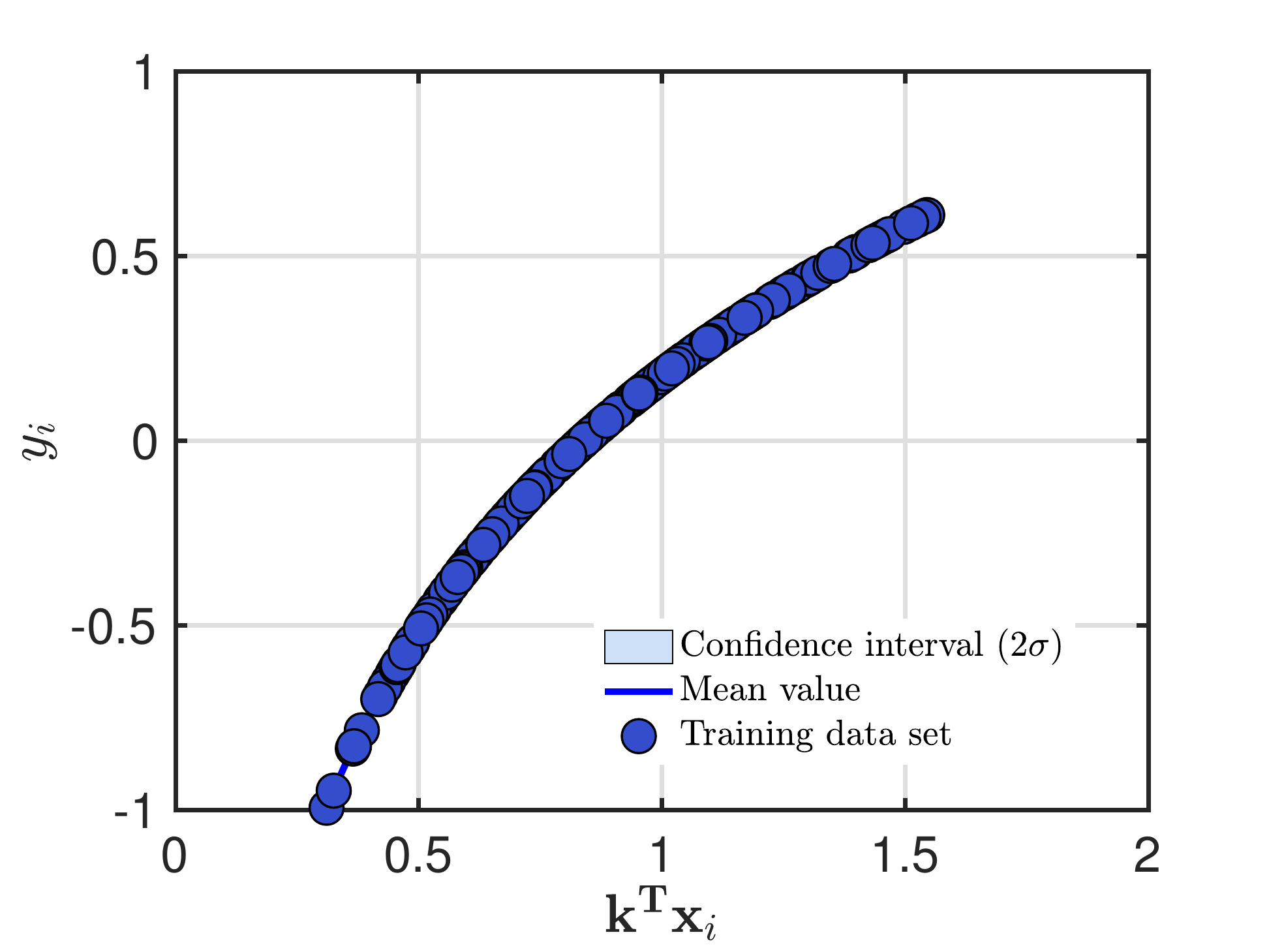}}
\subfigure[]{\includegraphics{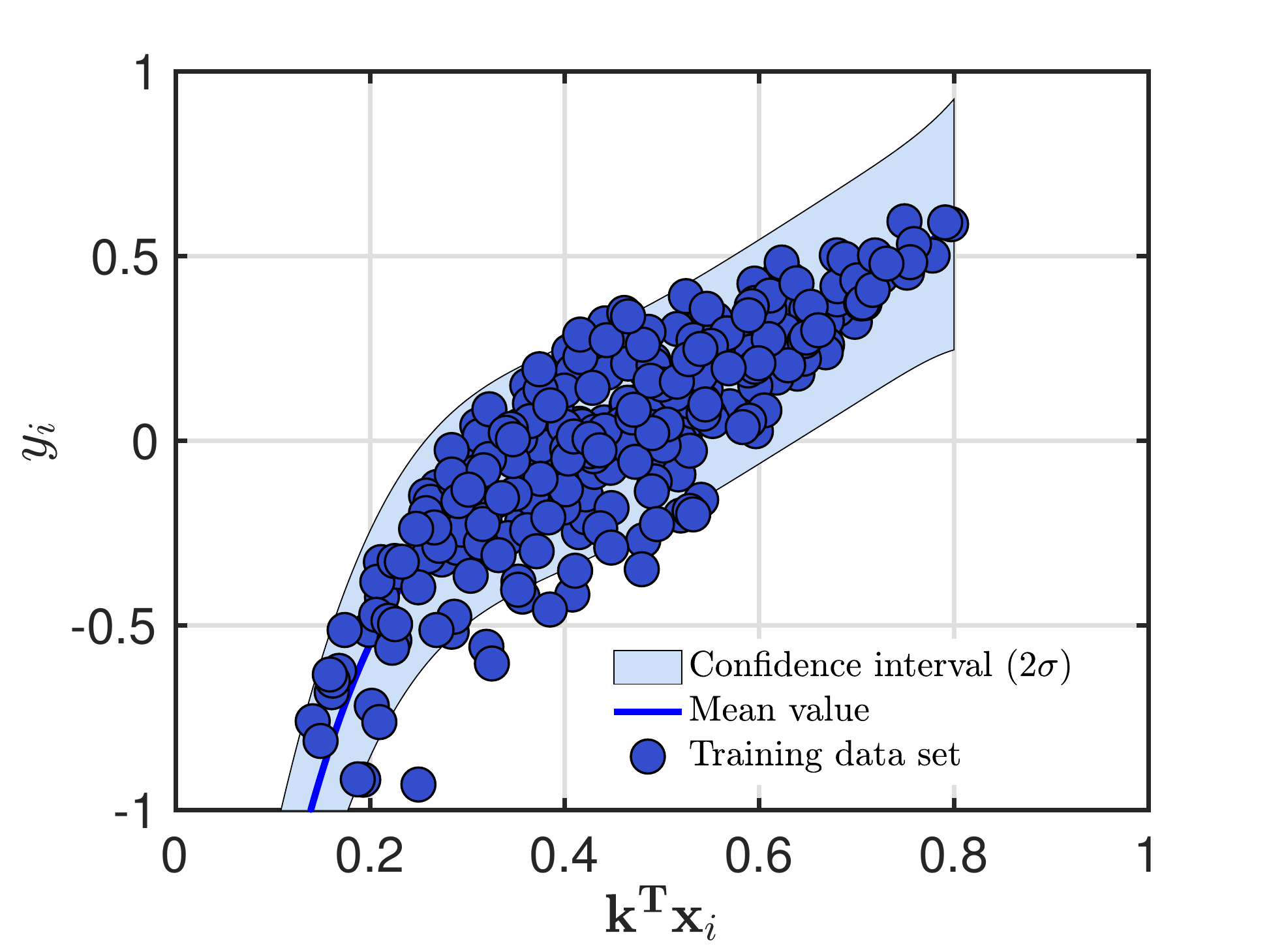}}
\end{subfigmatrix}
\caption{Sufficient summary plots of the function $y=log(x_1 + x_2 + x_3)$ where the horizontal axis is given by $\mathbf{k}^T (x_1, \; x_2, \; x_3)$ with a Gaussian process regression response surface. In (a) $\mathbf{k}=(1, 0, 0)^T$, in (b) $\mathbf{k}=(1/\sqrt{3}, 1/\sqrt{3}, 1/\sqrt{3})^T$, while in (c) $\mathbf{k}=(0.2, 0.5, 0.2)^T$.}
\label{figure_gp_motivation}
\end{figure}

We now return to the sufficient summary plots in Figure~\ref{figure_intro}. Assume we did not know the true function $f(x)=\text{log}(x_1 + x_2 + x_3)$ and that subfigures (a), (b) and (c) were simply the outcome of any one of the aforementioned algorithms. How do we determine that the subspace in (b) is ideally suited for the data? The logic we pursue in the paper is as follows. Assume we utilize Gaussian process regression (see Chapter 2 in~\cite{rasmussen2006gaussian}) to estimate the mean and standard deviation of the response $y_i$, as shown in Figure~\ref{figure_gp_motivation}. Then our criterion for selecting $\mathbf{k}$ in (b) is clear: it reduces the Gaussian process yielded standard deviation. In other words, for the same value of $\vx_i$ there is only one unique  $y_i$, ensuring that the deviation from the mean is near zero. 

There are two key ideas that emerge from this discussion. The first is that the posterior mean of a Gaussian process computed on a dimension-reducing subspace is a good candidate for $g$. In fact, this idea has been previously studied in \cite{tripathy2016gaussian} and \cite{liu2017dimension}, albeit using different techniques than those presented here. The second idea is that the suitability of a dimension-reducing subspace is reflected by the posterior variance: should the posterior variance be too large, as in Figure~\ref{figure_gp_motivation}(a), then one may need to opt for a different choice of $\vk$. 

In a nutshell, our objective in this paper is to offer an algorithm for computing a dimension-reducing subspace, under the constraint that $g$ is the posterior mean of a Gaussian process on this subspace. The remainder of this paper is structured as follows. In section \ref{sec:ridge} we survey the connections between ridge, active and sufficient dimension reduction subspaces. Then, motivated by the techniques in \cite{tripathy2016gaussian} and \cite{constantine2016anear}, we introduce an algorithm for computing a ridge subspace that uses tools from Gaussian process regression and manifold optimization. This is followed by numerical examples in section \ref{sec:results}.

%
%These two ideas motivate this paper. In section \ref{sec:sdr} we advance this idea further with four sufficient dimension reduction methods: SIR, SAVE CR and MAVE. Then, motivated by the work of~\cite{constantine2016anear}, we present a new algorithm for estimating the dimension-reducing subspace using manifold optimization (see~\ref{sec:manifold}). This is followed by ideas for computing the inverse map, estimating probabilities on the GP regression surface and computing individual parameter sensitivities. We close this paper with two case studies that show not just the efficacy of the proposed algorithm, but also the utility of the aforementioned ideas for the approximation theory, statistical regression and engineering design communities. 
%
%

\section{Ridge functions}
\label{sec:ridge}
If there exists a function $f: \mathbb{R}^{d} \rightarrow \mathbb{R}$ and a matrix $\mM \in \mathbb{R}^{d \times m}$, with $m \leq d$, that satisfy
\begin{equation}
f(\vx) = g(\mM^T \vx),
\label{ridge_def}
\end{equation}
for a function $g: \mathbb{R}^{m} \rightarrow \mathbb{R}$, then $f$ is a \emph{ridge function} \cite{pinkus2015ridge}. We call the subspace associated with the span of $\mM$ its \emph{ridge subspace}, denoted by $\mathcal{S}(\mM)$. We also assume that the columns of $\mM$ are orthonormal, i.e., $\mM^T \mM = \mI$. The above definitions imply that the gradient of $f$ is zero along directions that are orthogonal to $\mathcal{S}(\mM)$. In other words, if we replace $\vx$ with $\vx + \vh$ where $\mM^T \vh = 0$, then it is trivial to see that $f(\vx + \vh) = g(\mM^T (\vx + \vh)) = f(\vx)$. 
\subsection{Computing the optimal subspace}
So, given point evaluations of $f$, how does one compute $\mM$? Also, what are the properties of $\mM$? In their paper, Constantine et al.~\cite{constantine2016anear} draw our attention to the conditional and marginal densities defined on the subspace coordinates of the ridge subspace and its orthogonal complement. They present two main ideas. 

The first is the \emph{orthogonal invariance} associated with $\mM$, i.e., we are not interested in estimating a particular $\mM$, but rather the subspace of $\mM$---i.e., the ridge subspace. To study this further, we place a few additional assumptions on~(\ref{ridge_def}). Let the joint probability density function $\rho(\vx)$ and marginal densities $\rho_i$ be related by 
\begin{equation}
\bm{\rho}(\vx) = \prod_{i=1}^{d} \rho(x^{(d) } ).
\end{equation}
We assume that $f$ is square integrable with respect to this probability density function
\begin{equation}
\int_{\bm{\rho}} f^2(\vx) \bm{\rho} (\vx) d\vx < \infty,
\end{equation}
where the domain of $f$ is the support of $\bm{\rho}$. Define $\mathcal{S}(\mN)$ to be the orthogonal complement of the ridge subspace, where
\begin{align}
\mN = (\text{null}(\mM^T)) \; \; \; \text{with} \; \; \; \mN \in \mathbb{R}^{d \times \left(d-m \right) }. 
\end{align}
Placing the columns of $\mM$ and $\mN$ together in a single matrix
\begin{equation}
\mG=\left[\begin{array}{cc}
\mM & \mN\end{array}\right],
\end{equation} 
readily implies that $\mG \mG^{T} = \mI$. We now write the full decomposition of $f$ as
\begin{equation}
f\left(\vx\right)=f\left(\mG\mG^{T}\vx\right)=f\left(\mM\mM^{T}\vx+\mN\mN^{T} \vx\right)=f\left(\mM \vu+\mN \vv\right),
\end{equation}
with the following subspace coordinates
\begin{equation}
\vu \in \mathbb{R}^{m} \; \;  \text{and} \; \;  \vv \in \mathbb{R}^{d-m},\\
\end{equation}
where
\begin{equation}
\vu = \mM^{T} \vx \; \; \; \text{and} \; \; \;  \vv = \mN^{T} \vx. 
\label{def_of_subspace_coords}
\end{equation}
It should be noted here that if $f$ admits a ridge function structure, then the gradient of $f$ along the directions $\mN \vv$ are expected to be zero. 

For a fixed subspace coordinate $\vu$, Constantine et al. \cite{constantine2016anear} write the conditional expectation of $f$ to be
\begin{equation}
\mathbb{E}\left(f | \vu,\mM \right) =\int f\left( \mM \vu+\mN \vv       \right)\pi\left( \vv | \vu\right)d \vv 
\label{conditional_exp}
\end{equation}
where the conditional probability $\pi\left( \vv | \vu\right)$ can be expressed in terms of its marginal $\pi \left( \vu\right)$ and $\pi \left( \vu, \vv \right)$ joint probabilities
\begin{equation}
\pi\left( \vv| \vu\right)=\frac{\pi\left( \vu, \vv\right)}{\pi\left( \vu\right)}=\frac{\rho\left( \mM \vu+ \mN \vv\right)}{\int\rho\left(\mM \vu+\mN \vv\right)d \vv}.
\end{equation}
To demonstrate the orthogonal invariance associated with $\mM$ we replace it with $\mM \mQ$ where $\mQ \in \mathbb{R}^{m \times m}$ is an orthogonal matrix. Plugging this into~\eqref{conditional_exp} yields
\begin{align}
\mathbb{E}\left(f | \vu,  \mM \mQ \right) &=\int f\left(   \mM \mQ \left(\mM \mQ\right)^{T} \vx   +\mN \vv         \right) \pi\left( \vv | \vu\right)d \vv  \\
&=\int f\left(  \mM \underbrace{\mQ \mQ^T}_{\mI} \mM^T \vx +\mN \vv         \right)  \pi\left( \vv | \vu\right)d \vv  \\
&=\int f\left(  \mM \vu +\mN \vv         \right)  \pi\left( \vv | \vu\right)d \vv. 
\end{align}
It can be easily shown that the conditional probability also remains the same under the orthogonal transformation $\mM \mQ$. This demonstrates that the particular choice of $\mM$ does not matter, but rather its subspace does. In \cite{constantine2016anear}, the authors draw our attention to Theorem 8.3 in Pinkus \cite{pinkus2015ridge}, which proves that $\mathbb{E}\left(f|\vu,\mM \right)$ is the \emph{optimal ridge profile}---a function---in the $L_2$ norm. 

So, how do we compute this \emph{optimal subspace} associated with this ridge profile? The second point made in~\cite{constantine2016anear} is that this subspace can be computed by solving the following quadratic form
\begin{equation}
\begin{aligned}
& \underset{\mM}{\text{minimize} }
& &  \omega(\mM) =  \frac{1}{2} \int_{\bm{\rho}} \left( f   \left(\vx\right)-  \mathbb{E}\left( f | \vu,\mM \right) \right)^{2} \bm{\rho} \left(\vx \right) d\vx \\
& \text{subject to} 
& & \mM \in\mathbb{G}\left(m,d\right).
\end{aligned}
\label{solveme2}
\end{equation}
The constraint in~\eqref{solveme2} restricts $\mM$ to be from the subspace $\mathbb{G}(m, d)$, which denotes the space of $m$-dimensional subspaces of $\mathbb{R}^{d}$, i.e., the \emph{Grassman} manifold (see page 30 in \cite{absil2009optimization}). Recognizing~\eqref{solveme2} as a manifold optimization problem permits us to compute the gradients on the Grassman manifold. Following page 321 in~\cite{edelman1998geometry}, and denoting the gradient on the Grassman manifold by the symbol $\nabla_{\mathbb{G} }$, we have
\begin{align}
\nabla_{\mathbb{G} } \omega\left(\mM\right)&=\int_{\bm{\rho}  }\left(f\left( \vx\right)-\mathbb{E}\left(f|\vu,\mM  \right)\right)\left(- \nabla_{\mathbb{G}} \mathbb{E}\left(f|\vu,\mM \right)\right)\bm{\rho} \left( \vx\right)d \vx \\
&=\int_{\rho}\left(\mathbb{E}\left(f| \vu,\mM \right)-f\left(\vx\right)\right)\left(\mI-\mM \mM^{T} \right)\frac{\partial}{\partial \mM}\mathbb{E}  \left(f|\vu, \mM \right) \bm{\rho}\left( \vx\right)d \vx.
\label{equ_ridge_grads}
\end{align}
A few remarks regarding this optimization problem are in order. First, to compute derivative inside the integral, we require $f \in C^{1} \left( \mathbb{R}^{d} \right)$, which denotes a class of real-valued functions with continuous first derivatives. Provided the gradients $\frac{\partial}{\partial \mM}\mathbb{E}  \left(f|\vu, \mM \right) \in \mathbb{R}^{m \times d}$ can be determined, solutions to this optimization problem can be computed using an appropriate gradient-based optimizer. In subsection~\ref{sec:manopt} we use tools from Gaussian processes to estimate these gradients.  Second, \eqref{solveme2} is not necessarily convex, therefore we are not guaranteed a unique global minimum during an optimization. However, as we demonstrate in section \ref{sec:results}, near local optimal solutions can be computed. 

Another salient point concerns the choice of the manifold itself. In using the Grassman manifold, we enforce the assumption of \emph{homogenity}, i.e., $\omega \left( \mM \right) \equiv \omega \left(\mM \mQ \right)$ implying---as stated previously---that the specific entries in $\mM$ do not matter, but rather its subspace does. In other words, our objective is invariant to any choice of the basis \cite{edelman1998geometry}. In \ref{sec:gp} we discuss the ramifications of this assumption when using Gaussian process regression. 

\subsection{Connections between ridge subspaces and sufficient dimension reduction}
\label{approx_vs_reg}
One approach to estimate the ridge subspace is to develop a computational method that implements~\eqref{solveme2}. In subsection~\ref{sec:postvar} we detail such a procedure using Gaussian process regression. However, other algorithms that utilize ideas from sufficient dimension reduction theory can also be leveraged. The key is to understand the relationship between the subspace computed from techniques like SIR, SAVE and CR and the \emph{ridge subspace}. As a prelude to exploring these connections, it is important to distinguish statistical regression from approximation. 

In the regression setting we are given independent and identically distributed random variables $(\vx_i, y_i)$ from an \emph{unknown} joint distribution $\pi(\vx, y)$, where $i=1, \ldots, N$ for some $N$. Our objective is to characterize the conditional dependence of $y$ on $\vx$ either through its expectation $\mathbb{E}\left(y| \vx\right)$ or its variance $\sigma^2\left(y| \vx\right)$. This requires a model $y = h(\vx) + \epsilon$ for predicting $y$ given $\vx$, where $\epsilon$ is a zero-mean random error that is independent of $\vx$. This implies that for the same $\vx_{*}$ we have multiple values of $y_{*}$. Here $h$ is a parametric function and will typically be obtained by penalizing the errors in prediction via an appropriate \emph{loss function}, the most common one being the $l_2$ norm error \cite{friedman2001elements}. In the approximation setting, we are given input / output pairs $(\vx_i, y_i)$ obtained by evaluating a computer model under a known density $\pi(\vx)$ via an appropriately chosen \emph{design of experiment}. These inputs may be either design variables, boundary conditions or state variables. As the computer model, in general, does not have any random error associated with it, multiple evaluations for a particular $\vx_{*}$ will always yield the same $y_{*}$. In this context, our objective is to obtain a useful approximation $y = f(\vx)$ valid for \emph{all} $\vx$ over $\pi(\vx)$.  

Now, despite these differences, ideas from sufficient dimension reduction can be used for approximation. Glaws et al.~\cite{glaws2017inverse} detail the mathematical nuances involved in doing so for SIR and SAVE. They prove that that the conditional independence of the inputs and outputs in the approximation setting---i.e., for a deterministic function---is equivalent to the function being a ridge function (see Theorem 6 in~\cite{glaws2017inverse}). But is there any benefit in using SIR, SAVE and other sufficient dimension reduction methods for dimension reduction when approximating? Adragni and Cook \cite{Adragni2009} suggest that moment-based sufficient dimension reduction techniques such as SIR and SAVE, which are designed to provide estimates of the minimum sufficient linear reduction, are not tailored for prediction. They further suggest that model-based inverse regression techniques, such as those presented in \cite{Cook2007}, may be more useful. In section~\ref{sec:results} we test their suggestion by comparing the results of our proposed algorithm with SIR, SAVE, MAVE and CR.

%\begin{theorem} (Theorem 6 from~\cite{glaws2017inverse})
%Let $\left(\Omega,\varSigma,P\right)$ be a probability triplet with sample space $\Omega$, $\sigma$- algebra $\varSigma$ and probability measure $P$. Let $\mM \in \mathbb{R}^{m \times d}$ be a matrix. Define a function $f:\mathbb{R}^{d}\rightarrow\mathbb{R}$ that admits
%\begin{equation}
%y = f(\vx), \; \; \text{with} \; \; \vx \in \mathbb{R}^{d},
%\end{equation}
%where $\vx$ and $y$ are random variables. Then $y$ is linearly independent with respect to $\mM \vx$ if and only if $y=g(\mM \vx)$ where $g:\mathbb{R}^{m}\rightarrow\mathbb{R}$.
%\end{theorem}
%[From Glaws] Conditional independence of the inputs and output for a deterministic function is equivalent to the function being a ridge function. 

%
\subsection{Connections between ridge subspaces and active subspaces}
Active subspaces \cite{constantine2015active} refer to a set of ideas for parameter-based dimension reduction in the \emph{approximation} setting (see subsection~\ref{approx_vs_reg}). To compute the active subspace of a function $f$, we require that $f$ be differentiable and Lipschitz continuous---implying that the norm of its gradient $\nabla f_{\vx}$ is bounded. Then one assembles the covariance matrix $\mC$
\begin{equation}
\mC=\int\nabla f_{\vx} \nabla f_{\vx} \rho\left(\vx\right)d \vx,
\end{equation}
also known as the outer product of the gradient matrix\footnote{Note that this is a symmetric positive semidefinite matrix.}. What follows is an eigenvalue decomposition of $\mC$, where the first $m$ eigenvectors are used to define the \emph{active} subspaces and remaining $d-m$ eigenvectors the \emph{inactive} subspaces; the theorem below connects active subspaces and ridge subspaces.
\begin{theorem}
If we assume that $\mC$ admits the eigendecomposition
\begin{equation}
\mC=\left[\begin{array}{cc}
\mM & \mN\end{array}\right]\left[\begin{array}{cc}
\bm{\Lambda}_{1}\\
 & \bm{\Lambda}_{2}
\end{array}\right]\left[\begin{array}{cc}
\mM & \mN\end{array}\right]^{T} \; \; \; \text{where} \; \; \; \bm{\Lambda}_{1} \in \mathbb{R}^{m \times m},
\end{equation}
then 
\begin{equation}
\left(\int\left(f\left(\vx\right)-\mathbb{E}\left(f| \mM^{T} \vx\right)\right)^{2}\rho\left( \vx\right)d \vx \right)^{\frac{1}{2}}\leq  C\;\left( \text{trace} \left(\Lambda_{2}\right) \right)^{ \frac{1}{2}},
\label{paul_thm}
\end{equation}
where $C$ is a constant that depends on $\rho(\vx)$.
\end{theorem}
The proof follows from Theorem 4.4 in~\cite{constantine2015active}. It is important to emphasize that, in practice, the right-hand side of (\ref{paul_thm}) is difficult to compute because we do not have access to the eigenvalues associated with the covariance matrix that yields eigenvectors $[\mM, \mN]$. Furthermore, while the decay in the eigenvalues above can be used to set the size of $m$ when using active subspaces, when approximating ridge subspaces, one has to opt for other criteria---for instance, the authors in \cite{tripathy2016gaussian} utilize a Bayesian information criterion to determine $m$. 
%
%
%
%then 
%

\section{Gaussian processes}
\label{sec:gp}
The definition of a Gaussian process is a ``collection of random variables, any finite number of which have a joint Gaussian distribution" \cite{rasmussen2006gaussian}. It is completely defined by its mean and covariance functions. Restating the definition in the form we need, we have
\begin{equation}
f\left(\vx \right) \sim   \mathcal{GP} \left(\mu\left( \vu \right), \bm{k}\left(\vu \right)\right),
\label{equ_GPR}
\end{equation}
where $\mu$ is the mean function and $\bm{k}$ represents a parameterized covariance matrix, albeit with a slight abuse of notation. It is important to note that we define our Gaussian process on the reduced coordinates $\vu = \mM^T \vx$. As a result, the computation of the Gaussian process itself does not become prohibitive. 

In this section we have two main goals. First, to define a Gaussian ridge function and to outline its properties. Second, to provide an algorithm that computes a Gaussian ridge function. To ease our exposition, we assume the existence of a set of:
\begin{itemize}
\item A set of input/output training data pairs:
\begin{align}
\begin{split}
\left( \hat{\vx}_i, \hat{f}_{i} \right), \; \; \; \; \text{for} \; i=1, \ldots, N_{train}, \\
\hat{\vf}=\left(\begin{array}{c}
\hat{f}_{1}\\
\vdots\\
\hat{f}_{N_{train}}
\end{array}\right), \; \; \; \; \hat{\mU}=\left(\begin{array}{ccc}
\hat{\vu}_{1}, & \ldots, & \hat{\vu}_{N}\end{array}\right)^{T} \; \; \; \; \text{and} \; \; \; \; \hat{\vu}_{i}=\mM^{T}\hat{\vx}_{i}.
\end{split}
\end{align}
\item A set of input/output testing data pairs:
\begin{equation}
\begin{split}
\left( \tilde{\vx}_i, \tilde{f}_{i} \right), \; \; \; \; \text{for} \; i=1, \ldots, N_{test}, \\
\tilde{\vf}=\left(\begin{array}{c}
\tilde{f}_{1}\\
\vdots\\
\tilde{f}_{N_{test}}
\end{array}\right), \; \; \; \; \tilde{\mU}=\left(\begin{array}{ccc}
\tilde{\vu}_{1}, & \ldots, & \tilde{\vu}_{N}\end{array}\right)^{T} \; \; \; \; \text{and} \; \; \; \; \tilde{\vu}_{i}=\mM^{T}\tilde{\vx}_{i}.
\end{split}
\end{equation}
\end{itemize}

%
% $N_{train}$ input/output \emph{training data} pairs $(\hat{\vx}_{i}, \hat{f}_i)$ for $i=1, \ldots, N_{train}$; the superscript $\left( \hat{\cdot} \right) $ is used for identifying the training data pairs. We then define the following quantities
%\begin{equation}
%\hat{\vf}=\left(\begin{array}{c}
%\hat{f}_{1}\\
%\vdots\\
%\hat{f}_{N}
%\end{array}\right), \; \; \; \; \hat{\mU}=\left(\begin{array}{ccc}
%\hat{\vu}_{1}, & \ldots, & \hat{\vu}_{N}\end{array}\right)^{T} \; \; \; \; \text{and} \; \; \; \; \hat{\vu}_{i}=\mM^{T}\hat{\vx}_{i}.
%\label{notation_1}
%\end{equation}
%\noindent For convenience, we also define 
%\begin{equation}
%\mU_{\ast} = \left(\begin{array}{ccc}
%\tilde{\vu}-\hat{\vu}_{1}, & \ldots, & \tilde{\vu}-\hat{\vu}_{N}\end{array}\right)^{T},
%\end{equation}
%where the variable $\tilde{\vu}$ represents a generic test point.
\subsection{Gaussian ridge functions (posterior mean)}
Succinctly stated, a Gaussian ridge function is the posterior mean of the Gaussian process in~\eqref{equ_GPR}, written as
\begin{equation}
\bar{g}\left(\vu \right) = \bm{k}\left(\vu, \hat{\mU} \right)\bm{ \beta}.
\label{equ_posterior}
\end{equation}
It is a linear sum of kernel functions with coefficients $\bm{\beta}$---an outcome of the representer theorem (see page 132 of~\cite{rasmussen2006gaussian}). Before we define these coefficients, we first focus on the covariance term on the right-hand side in~\eqref{equ_posterior}. In what follows, we employ the squared exponential covariance kernel
\begin{equation}
k\left(\vu_{i},\vu_{j}\right)=\sigma_{f}^{2}\; \text{exp} \left(-\frac{1}{2}\left(\vu_{i}-\vu_{j}\right)^{T}\bm{\Gamma}^{-1}\left(\vu_{i}-\vu_{j}\right)\right),
\end{equation}
where the diagonal matrix $\bm{\Gamma}$ is given by
\begin{equation}
\bm{\Gamma}=\left[\begin{array}{ccc}
l_{1}^{2} &  & 0\\
 & \ddots\\
0 &  & l_{m}^{2}
\end{array}\right].
\label{equ:gaussian_diagonal}
\end{equation}
The constant $\sigma^2_f$ is known as the \emph{signal variance} and constants $l_1, \ldots, l_m$ are the \emph{correlation lengths} along each of the $m$ coordinates of $\vu$. The parameterized covariance matrix $\bm{k}\left(\vu, \hat{\mU}  \right) \in \mathbb{R}^{1 \times N}$ in \eqref{equ_posterior} can now be written as
\begin{equation}
\bm{k}\left( \vu, \hat{\mU}   \right)=\left( \begin{array}{ccc}
k\left(\vu, \hat{\vu}_{1}    \right), & \ldots, & k\left(\vu ,\hat{\vu}_{N} \right)\end{array}\right).
\end{equation}
The vector of coefficients $\bm{\beta} \in \mathbb{R}^{N}$ is given by
\begin{equation}
\bm{\beta}=\left( \mK+\sigma_{n} \mI\right)^{-1} \hat{\vf},
\end{equation}
where the $(i,j)$-th entry of the matrix $\mK \in \mathbb{R}^{N \times N}$ is given by
\begin{equation}
\mK(i,j)=k\left(\hat{\vu}_{i} ,  \hat{\vu}_{j}  \right),
\end{equation}
in other words it depends only on the training data set. Once again, for notational convenience, we define the posterior mean values at the training and testing points by
\begin{equation}
\bar{\vg}_{train}=\left(\begin{array}{c}
\bar{g}\left(\hat{\vu}_{1}\right)\\
\vdots\\
\bar{g}\left(\hat{\vu}_{N_{train}} \right)
\end{array}\right), \; \; \; \; \; \bar{\vg}_{test}=\left(\begin{array}{c}
\bar{g}\left(\tilde{\vu}_{1}\right)\\
\vdots\\
\bar{g}\left(\tilde{\vu}_{N_{test}} \right)
\end{array}\right).
\end{equation}

To aid the optimization strategy we describe in the forthcoming section, we are interested in obtaining the gradients of \eqref{equ_posterior}; this yields
\begin{equation}
\begin{aligned}
& & \frac{\partial \bar{g}}{\partial \vu} & =\frac{\partial \bm{k} \left(\vu , \hat{\mU} \right)}{\partial \vu } \bm{\beta} \\
& & & = -\bm{\Gamma}^{-1}\tilde{\mU}\left(\bm{k} \left(\vu, \hat{\mU} \right)^{T}\odot \bm{\beta} \right),
\end{aligned}
\label{gp_grads}
\end{equation}
where the symbol $\odot$ indicates an element-wise product. In practice, any differentiable kernel may be used to compute the gradient of the Gaussian ridge function. Now, in using a Gaussian ridge function, we make the assertion that it is approximately the expectation of $f$ projected on $\vu = \mM^T \vx$, i.e.,
\begin{equation}
\mathbb{E}\left(f| \vu, \mM\right) \approx \bar{g}\left(\vu\right).
\label{approx_quality}
\end{equation}
Its gradients, can then be computed via
\begin{equation}
\begin{aligned}
\frac{\partial}{\partial \mM} \bar{g}\left(\vu\right) & = \frac{\partial \bar{g}}{\partial \vu}\frac{\partial\left(\mM^{T} \vx\right)}{\partial \mM} \\
& = -\bm{\Gamma}^{-1}\tilde{\mU}\left(\bm{k} \left(\vu, \hat{\mU} \right)^{T}\odot \bm{\beta} \right) \vx.
\label{equ:grads_st}
\end{aligned}
\end{equation}
It is clear that the quality of our approximation in~\eqref{approx_quality} is contingent upon a suitable choice of the hyperparameters
\begin{equation}
\bm{\theta}=\left\{ \sigma_{f}^{2},\sigma_{n}^{2},l_{1},\ldots,l_{m}\right\} .
\end{equation}
A well-worn heuristic to estimate these hyperparameters, which in itself is a non-convex optimization problem, is to maximize the marginal likelihood (see page 113~\cite{rasmussen2006gaussian})
\begin{equation}
\begin{aligned}
& \underset{\bm{\theta}}{ \text{maximize} } 
& &  \text{log} \; p\left(\vu, \bm{\theta}\right)  \\
& \text{subject to} 
& & 
\text{log} \; p\left(\vu, \bm{\theta}\right) =-\frac{1}{2} \vf^{T}\left( \mK+\sigma_{n}^{2} \mI\right)^{-1} \vf-\frac{1}{2} \text{log} \left| \mK+\sigma_{n}^{2} \mI \right|-\frac{N}{2} \text{log} \left(2\pi\right).
\end{aligned}
\label{gp_hyperparameters}
\end{equation}
The first term in $\text{log} \; p\left(\vu, \bm{\theta}\right)$, which depends solely on the covariance matrix and values of $f$, is the \emph{data-fit} term. The second term, which depends only upon the covariance function, is known as the \emph{complexity penalty}, while the third term is simply a normalizing constant. The problem of computing the maximum in~\eqref{gp_hyperparameters} can be readily solved via a gradient-based optimizer; formulas for the gradients of $\text{log} \; p$ with respect to the hyperparameters are given in section 5.9 of \cite{rasmussen2006gaussian}.

\subsection{The utility of the posterior variance}
\label{sec:postvar}
In our presentation of Gaussian ridge functions so far we have not discussed the posterior variance. It is given by 
\begin{equation}
\mathbb{V}\left[ \mathcal{GP} \right]= k\left(\vu, \vu\right)-k\left( \vu,\hat{\mU}\right)^{T}\left(\mK+\sigma_{n}^{2} \mI\right)^{-1} \vk\left( \vu,\hat{\mU}\right),
\end{equation}
and can be very useful in quantifying the suitability of a particular choice of $\mM$. Recall the plots in Figures~\ref{figure_intro} and \ref{figure_gp_motivation}. Our perception of a successful dimension reduction strategy through a sufficient summary plot is intimately tied to the variance in $f$ for a fixed coordinate in the dimension-reducing subspace. Here, we argue that the expectation of the posterior variance, $\mathbb{E}\left[\mathbb{V}\left(g\right)\right]$, can be used to gauge the overall effectiveness of the subspace approximant $\hat{\mM}$. 

We make the above argument for two reasons. First, different dimension reduction methods---tailored for the same projection $\mathbb{R}^{d} \rightarrow \mathbb{R}^{m}$---will likely yield different subspaces. In~\cite{cook2005sufficient} a clear distinction is made between dimension reduction techniques that identify a \emph{central subspace} from those that find a \emph{central mean subspace}; an algorithm for computing the \emph{central variance subspace} is introduced in~\cite{zhu2009dimension}. Our hypothesis is that regardless of the subspace sought, to infer the suitability of a particular $\hat{\mM}$ as a ridge subspace we must be able to compare different methods that by definition will yield different subspaces. To this end, while a metric based on the subspace angle $\phi$ (see page 329 of~\cite{golub2012matrix})---i.e.,
\begin{equation}
\begin{aligned}
\text{dist}\left( \mM,\tilde{\mM}\right) & =\left\Vert \mM\mM^{T}-\tilde{\mM}\tilde{\mM}^{T}\right\Vert _{2} \\
& = sin(\phi)
\end{aligned}
\end{equation}
---is useful for comparing the \emph{distance} between the \emph{ridge subspace} $\mM$ and its approximant $\tilde{\mM}$, it is limited to comparisons with one technique / subspace. The second point we articulate concerns cases when $m \geq 3$, where it is difficult to visually perceive how suitable a choice of $\bar{g}$ is. Here, the measure $\mathbb{E}\left[\mathbb{V}\left(g\right)\right]$ can be used (i) to quantify the error in the approximation, and (ii) to guide further computer simulations. 

\subsection{Computing the Gaussian ridge via manifold optimization}
\label{sec:manopt}
Recall the definition of the correlation lengths associated with the Gaussian process regression model in \eqref{equ:gaussian_diagonal}. If we were to set the correlation lengths to be the same, i.e., $l_1=l_2= \ldots = l_m$, then we satisfy our requirement of homogenity, and can optimize \eqref{solveme2} using the gradient computation in \eqref{equ_ridge_grads}. While this may be a potential solution strategy it does require us to modify the maximum likelihood criterion in \eqref{gp_hyperparameters}. To circumvent this, we propose to alter the constraint in \eqref{solveme2}; instead of optimizing over the Grassman manifold, we optimize over the Stiefel manifold,
\begin{equation}
\begin{aligned}
& \underset{\mM}{\text{minimize} }
& &  \omega(\mM) =  \frac{1}{2} \int_{\bm{\rho}} \left( f   \left(\vx\right)-  \bar{g} \left( \mM^{T} \vx  \right) \right)^{2} \bm{\rho} \left(\vx \right) d\vx \\
& \text{subject to} 
& & \mM \in\mathbb{S}t \left(m,d\right).
\end{aligned}
\label{solvemenew}
\end{equation}
Here $\mathbb{S}t \left(m,d\right)$ is the set of all $d \times m$ orthogonal matrices 
\begin{equation}
\mathbb{S}t\coloneqq\left\{ \mM\in\mathbb{R}^{d\times m}: \mM^{T} \mM= \mI_{m}\right\};
\end{equation}
it is an embedded submanifold of $\mathbb{R}^{d \times m}$ and is compact \cite{absil2009optimization}. Denoting the gradients on the Stiefel manifold by $\tilde{\nabla}$, gradients of $\omega$ can be computed via
\begin{align}
\nabla_{\mathbb{S}t} \omega\left(\mM\right)&=\int_{\bm{\rho}  }\left(f\left( \vx\right)-  \bar{g} \left( \mM^{T} \vx  \right)   \right)    \left(-\bar{\nabla}_{\mathbb{S}t}     \bar{g} \left( \mM^{T} \vx  \right)  \right)\bm{\rho} \left( \vx\right)d \vx \\
&=\int_{\rho}\left(    \bar{g} \left( \mM^{T} \vx  \right)     -f\left(\vx\right)\right)   \left(   \frac{\partial}{\partial \mM} \bar{g} \left( \mM^{T} \vx  \right) - \mM  \frac{\partial}{\partial \mM}  \bar{g} \left( \mM^{T} \vx  \right)^{T} \mM    \right) \bm{\rho}\left( \vx\right)d \vx,
\label{equ_ridge_grads_new}
\end{align}
(see 2.53 of \cite{edelman1998geometry}). We use this expression in our algorithm for estimating $\mM$ using a Gaussian ridge function. Ours is an iterative approach---optimizing over the Stiefel manifold for estimating $\mM$ and then over the space of the hyperparameters for determining $\bm{\theta}$, where the correlation lengths along each of the $m$ directions can be different. We remark here that our approach is similar to the alternating approach from Algorithm 2 in Constantine et al. \cite{constantine2016anear} that uses polynomial approximations. 

Our algorithmic workflow is given as follows. First, the data set must be split into training and testing data sets, where the former has $N_{train}$ input/output pairs while the latter has $N_{test}$ pairs. It is difficult to provide heuristics on what values to set for $N_{test}$ and $N_{train}$, however, we do require that both testing and training sets are distributed with respect to the measure $\rho(\vx)$.  

For our algorithm, we require an initial choice of the ridge $\mM$, which is readily computed by applying a QR factorization on a random normal matrix and then taking the first $m$ columns of $\mQ$ (see lines 3--5). This trick ensures that the columns of $\mM$ are orthogonal. It also implies that each call of our algorithm will yield a different result, by virtue of the fact that the initial starting point---i.e., $\mM_{0}$---is different. Following this, it is easy to project the $d$-dimensional data $\vx$ to the $m$-dimensional space (line 7).

\begin{algorithm}[H]
\label{algo}
\caption{Gaussian ridge function approximation.}
\begin{algorithm2e}[H]
\KwData{Input/output pairs $(\vx_j, f_j)$ where $j=1, \ldots, N_{train}+N_{test}$}
\KwResult{Matrix $\mM$ }
Split the data into training data set $(\hat{\vx}_i, \hat{f}_i ) \; \; \text{where} \; \;   i=1, \ldots, N_{train}$. \\
Assemble the testing data set $(\tilde{\vx}_i, \tilde{f}_i)\; \; \text{where} \; \;   i=1, \ldots, N_{test}$.\\
Initialize a random normal matrix $\mA \in \mathbb{R}^{d \times m}$.\\
Compute its QR factorization $\mA=\left(\begin{array}{cc} \mQ_{1} & \mQ_{2}\end{array}\right) \mR$, where $\mQ_{1} \in \mathbb{R}^{d \times m}$ and $\mM \coloneqq \mQ_{1}   $.\\
 \While{$\left|\triangle r\right|\leq\epsilon$}{
  Compute:
  \[
  \hat{ \vu}_{i} = \mM^T \hat{\vx}_{i} \; \; \;  \text{and} \; \; \; \tilde{\vu}_{i} = \mM^T \tilde{\vx}_{i}.
  \]\\
Solve the optimization problem using the training data $\left( \hat{ \vu}_{i}, \hat{\vf}_{i} \right)$:
\[
\bm{\theta}_{\ast} = \text{argmax}  \; \text{log} \; p\left(\hat{\vu}, \bm{\theta}\right).
\] \\
Solve the Stiefel manifold optimization problem using $\bm{\theta}_{*}$ on the testing data:
\[
\mM^{\ast} =\text{argmin} \; \frac{1}{2N_{test}}\left\Vert \tilde{\vf}-  \tilde{\vg}\left( \mM, \bm{\theta}_{\ast} \right)  \right\Vert _{2}^{2}.
\] \\
 Set: $\mM = \mM^{\ast}$.
 }
\end{algorithm2e}
\end{algorithm}

Next, we train a Gaussian process model using $\left( \hat{\vu}_{i}, \hat{\vf}_{i}  \right)$ and solve the optimization problem in~\eqref{gp_hyperparameters} (line 7) to obtain the hyperparameters $\bm{\theta}$. These values are then fed back into the Gaussian process posterior mean---i.e., our approximation of $\mathbb{E}\left(f| \vu, \mM\right)$---for prediction. In line 8, we optimize over the Stiefel manifold using the hyperparameters $\bm{\theta}$. This requires discretizing the integral in \eqref{solvemenew}, resulting in
\begin{align}
\begin{split}
r  & = \frac{1}{2 N_{test} } \sum_{i=1}^{N_{test}}\left(\tilde{f}_{i}- \bar{g} \left(\mM^{T}\tilde{\vx}_{i},  \boldsymbol{\theta}_{\ast} \right)\right)^{2} \\
 &  =  \frac{1}{2 N_{test} }  \left\Vert \tilde{\vf}-  \tilde{\vg}  \right\Vert _{2}^{2}.
\end{split}
\end{align}
The gradients of this function on the Stiefel manifold can then be obtained using
\begin{equation}
\nabla_{\mathbb{S}t} r \left( \mM \right) = \frac{1}{N_{test}}\sum_{i=1}^{N_{test}}\left(\tilde{f}_{i}- \bar{g} \left( \tilde{\vu}_{i},\theta\right)\right)\left\{ \frac{\partial \bar{g} \left( \tilde{\vu}_{i}\right)}{\partial \mM}- \mM \left(\frac{\partial \bar{g} \left( \tilde{\vu}_{i}\right)}{\partial \mM}\right)^{T} \mM\right\}, 
\end{equation}
where $\partial \bar{g} \left( \tilde{\vu}_{i} \right) /  \partial \mM $ is determined by plugging in the testing data into \eqref{equ:grads_st}. Both the values of $r$ and $\nabla_{\mathbb{S}t} r  \in \mathbb{R}^{d \times m}$ are provided to the manifold optimizer in step 8. This process is then repeated till the difference in residuals $\left|\triangle r\right|$ between successive iterations is below a chosen value $\epsilon$. 

\section{Numerical studies}
\label{sec:results}
Our codes for Algorithm 1 can be found at \url{https://www.github.com/~psesh/Gaussian-Ridges}. The codes are in \texttt{MATLAB} and require the \texttt{gpml} \cite{rasmussen2010gaussian} toolbox for Gaussian processes and \texttt{manopt}  \cite{manopt}, a set of utilities for manifold optimization. For the manifold optimization subroutine in our codes, we use the preconditioned Riemannian conjugate gradients algorithm of Boumal and Absi (see page 224 of \cite{boumal2015low}). For optimizing the hyperparameters in our Gaussian process models, we use the in-built (within \texttt{gpml}) conjugate gradient approach based on the Polak-Ribi\`{e}re method (see~\cite{polak2012optimization}). Initial values for all our hyperparameters were set to 0.

\subsection{An analytical linear ridge}
In this section, we study the outcome of the algorithm provided in subsection~\ref{sec:manopt} for discovering the ridge subspace associated with the simple problem
\begin{equation}
\begin{split}
f(x) = g\left( \mM^{T} \vx \right) \; \; \; \text{where} \; \; \; \mM = \left[\begin{array}{cc}
\vm_{1} & \vm_{2}\end{array}\right],  \\
\mM \in \mathbb{R}^{d \times 2}, \; \; \vx \in \mathbb{R}^{d} \; \; \; \text{and} \; \; g: \mathbb{R}^{2} \rightarrow \mathbb{R},
\end{split}
\label{example_gaussian}
\end{equation}
with  $\vx \in [-1, 1]^{d}$.  Here $g$ is a bivariate function given by
\begin{equation}
g \left(y_{1}, y_{2} \right) = y_{1} + y_{2}.
\end{equation}
We explore the result of the conjugate gradient optimizer by varying the number of training and testing points along with the number of dimensions $d$. As our optimization problem is non-convex, we expect different outcomes based on the initial value provided to the optimizer. As a result, we run each optimization numerous times---using the same training and testing data-set. In other words, each optimization trial begins with a random orthogonal matrix (see Step 3 in Algorithm 1) and then aims to minimize our objective function (see Step 8 in Algorithm 1). In the graphs below, we denote our optimized solution as
\begin{equation}
\mM^{\ast} = \left[\begin{array}{cc}
\vm_{1}^{\ast} & \vm_{2}^{\ast} \end{array}\right], \; \; \; \text{where} \; \; \; \mM^{\ast} \in \mathbb{R}^{d \times 2}.
\end{equation}

\subsubsection{Case $d=10$}
\begin{figure}
\begin{subfigmatrix}{2}% number of columns
\subfigure[]{\includegraphics{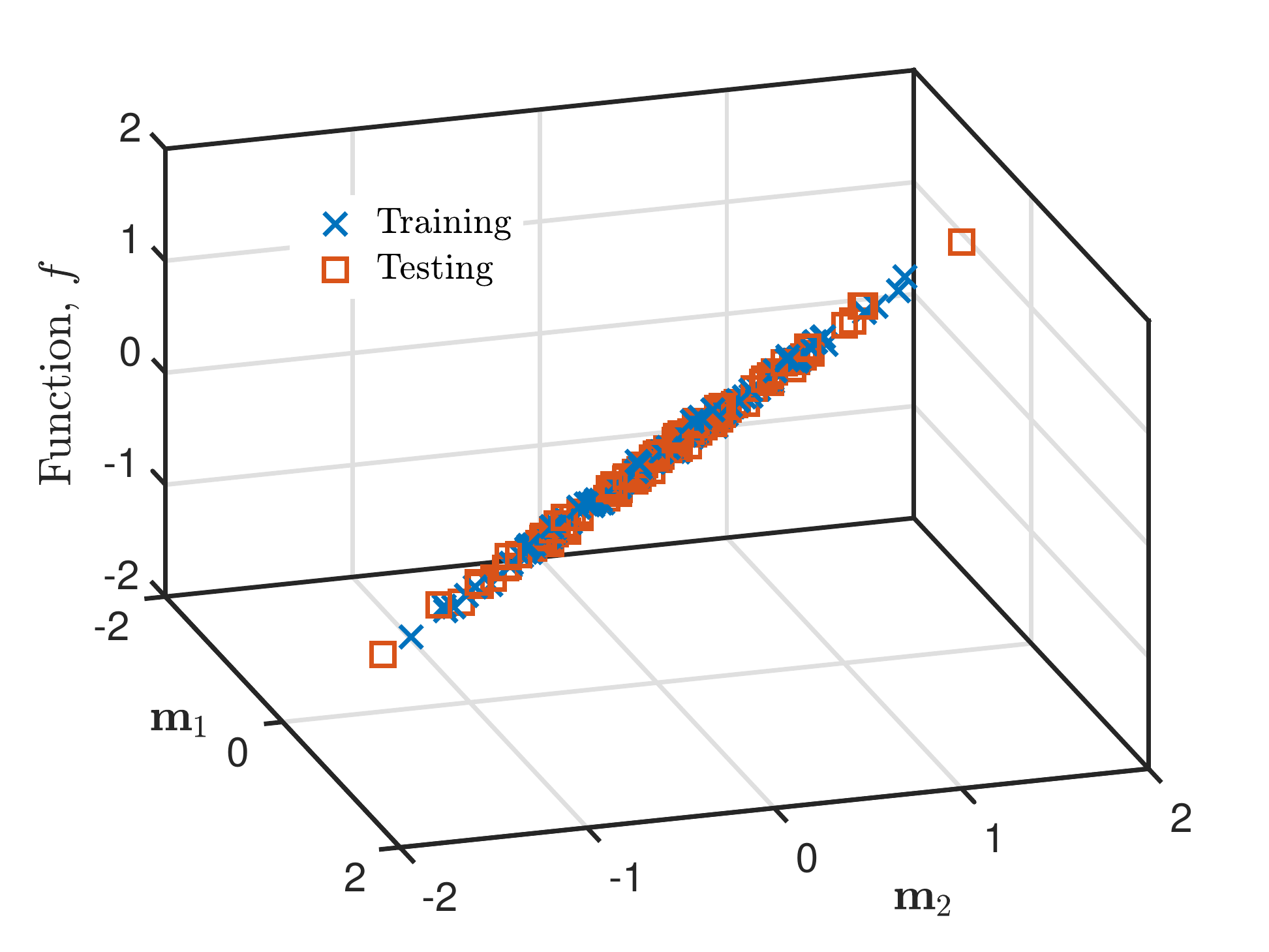}}
\subfigure[]{\includegraphics{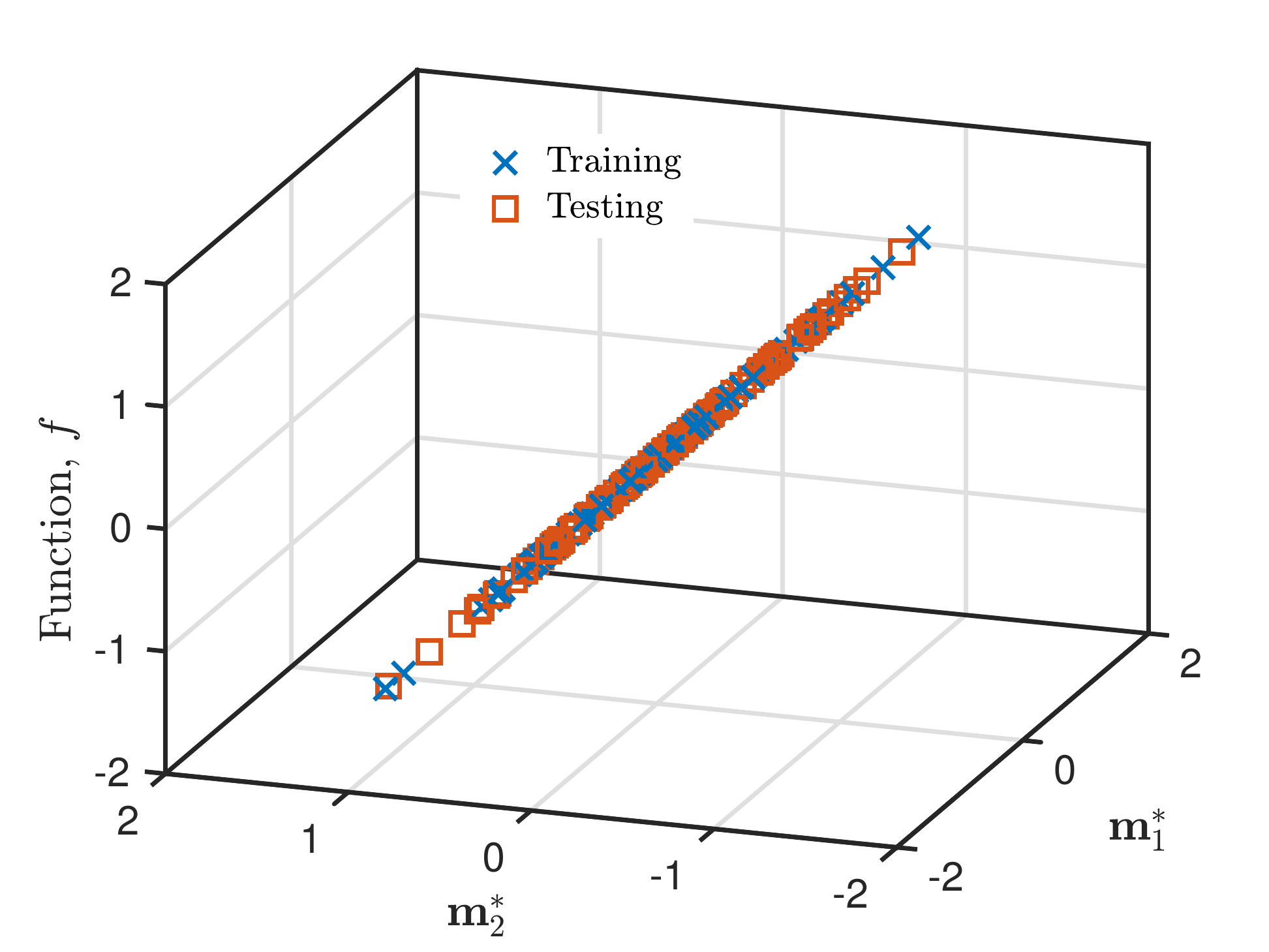}}
\subfigure[]{\includegraphics{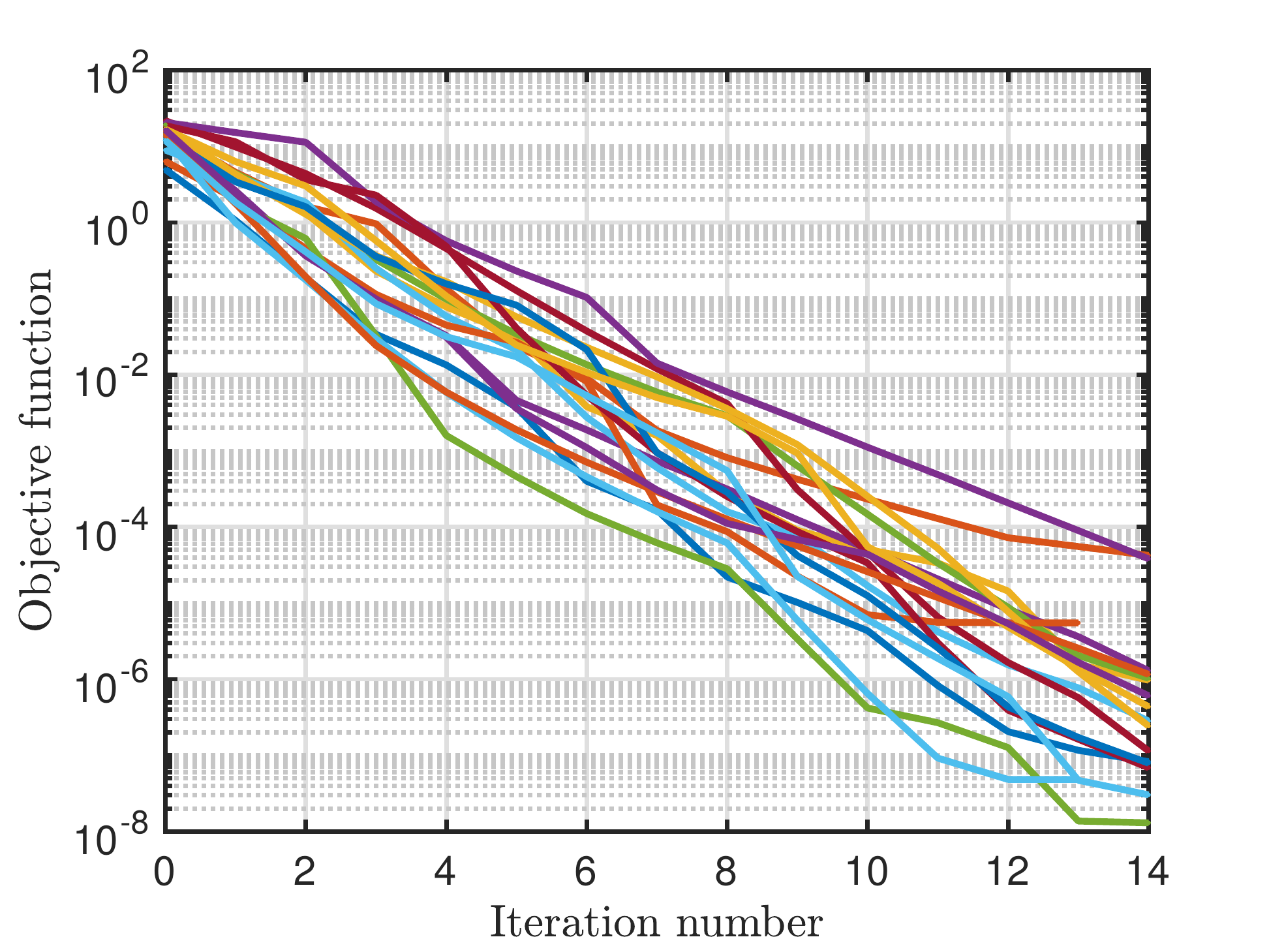}}
\subfigure[]{\includegraphics{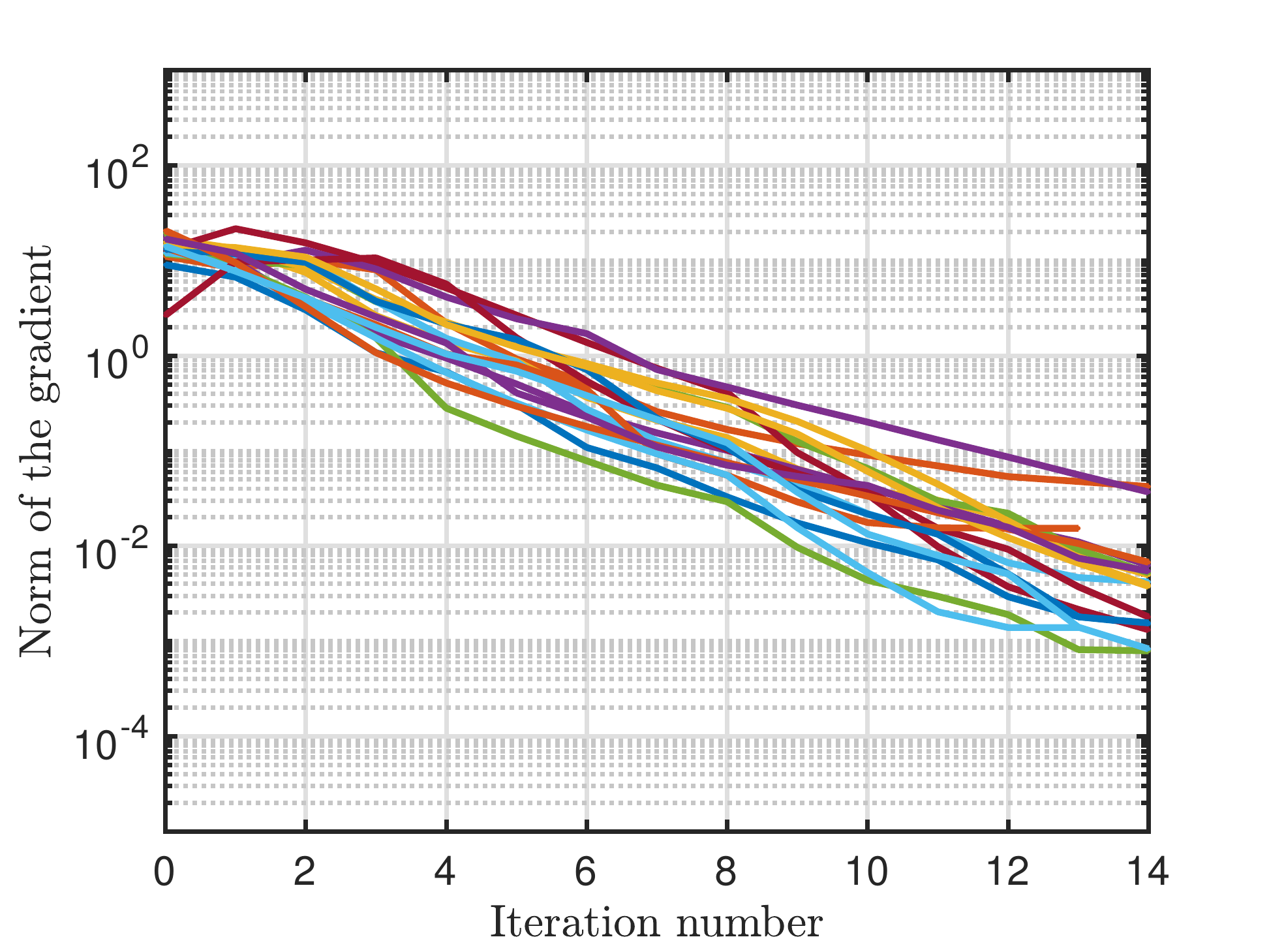}}
\end{subfigmatrix}
\caption{Results of the analytical linear ridge problem with $d=10$, $N_{test}=100$ and $N_{train}=100$ with 20 repetitions: (a) Sufficient summary plot on $\mM$; (b) sufficient summary plot on $\mM^{\ast}$; (c) objective function vs. iterations; (d) gradient norm vs. iterations.}
\label{gp_10}
\end{figure}

To begin, consider the isolated case where we set $N_{test} = 50$ and $N_{train} = 50$. Figure~\ref{gp_10}(a) illustrates the sufficient summary plot on $\mM$, while in (b) we plot the sufficient summary plot on the optimizer-yielded $\mM^{\ast}$, taken from a trial that yielded the lowest value of the objective function; clearly our algorithm offers a suitable approximation to this ridge function. We repeat this experiment 20 times, each time specifying a different initial orthogonal matrix. In Figure~\ref{gp_10}(c) we plot the change in the objective function over the number of iterations and in (d) we plot the norm of the gradient per iteration.

\begin{figure}
\begin{subfigmatrix}{2}% number of columns
\subfigure[]{\includegraphics{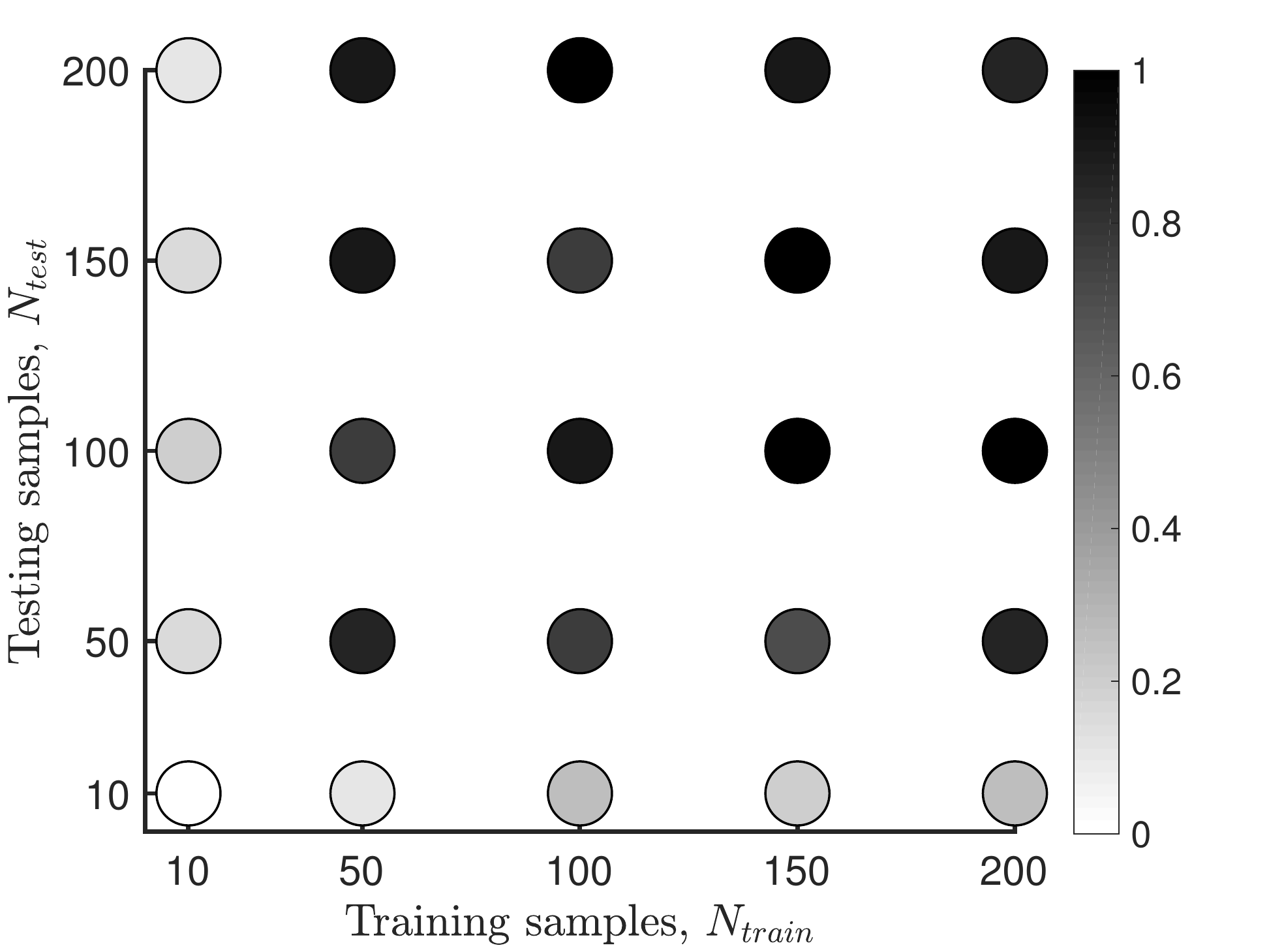}}
\subfigure[]{\includegraphics{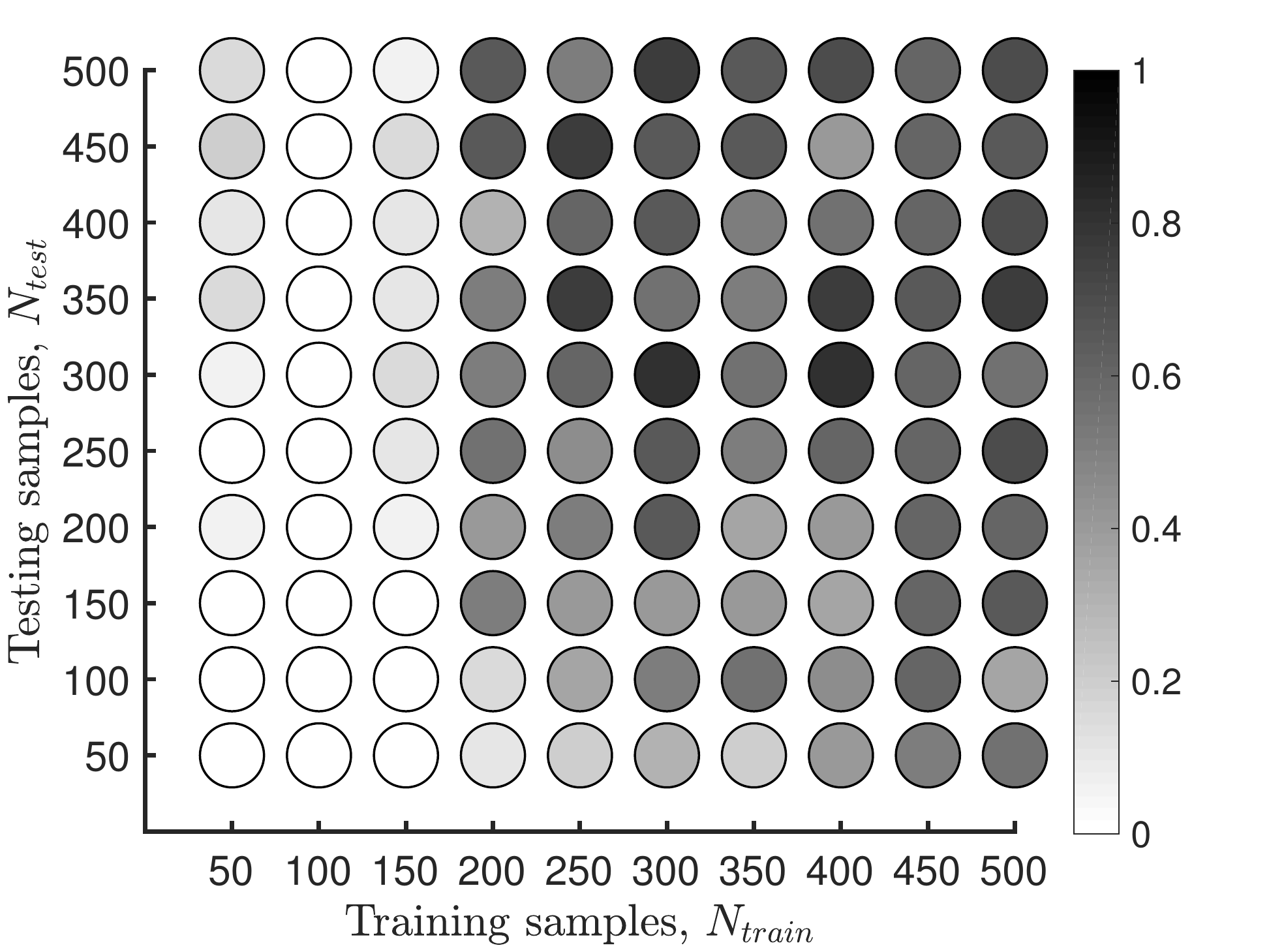}}
\end{subfigmatrix}
\caption{Average probability of success values from 20 trials, for different training and testing samples for the case where (a) $d=10$; (b) $d=100$ for the analytical linear ridge problem. Darker colored circles denote a greater probability of success.}
\label{gp_compare}
\end{figure}

As alluded to previously, it is difficult to offer bounds on the number of testing and training samples one should use in Algorithm~\ref{algo}. In Figure~\ref{gp_10}(d) we plot the probability of success for different combination of testing and training samples for this problem. We define an optimization trial to be successful if the objective function attains a value $\leq 0.005$. Furthermore, as the outcome of any optimization is dependent upon the initial point, we repeat each experiment with 20 trials and report the average values in the individual circles in Figure~\ref{gp_compare}(a). Circles that are black have, on average, a high chance of recovering the ridge subspace. For this example problem, it is clear that beyond $N_{test} \geq 50$ and $N_{train} \geq 50$, ridge recovery is obtained. 

\subsubsection{Case $d=100$}
\begin{figure}
\begin{subfigmatrix}{2}% number of columns
\subfigure[]{\includegraphics{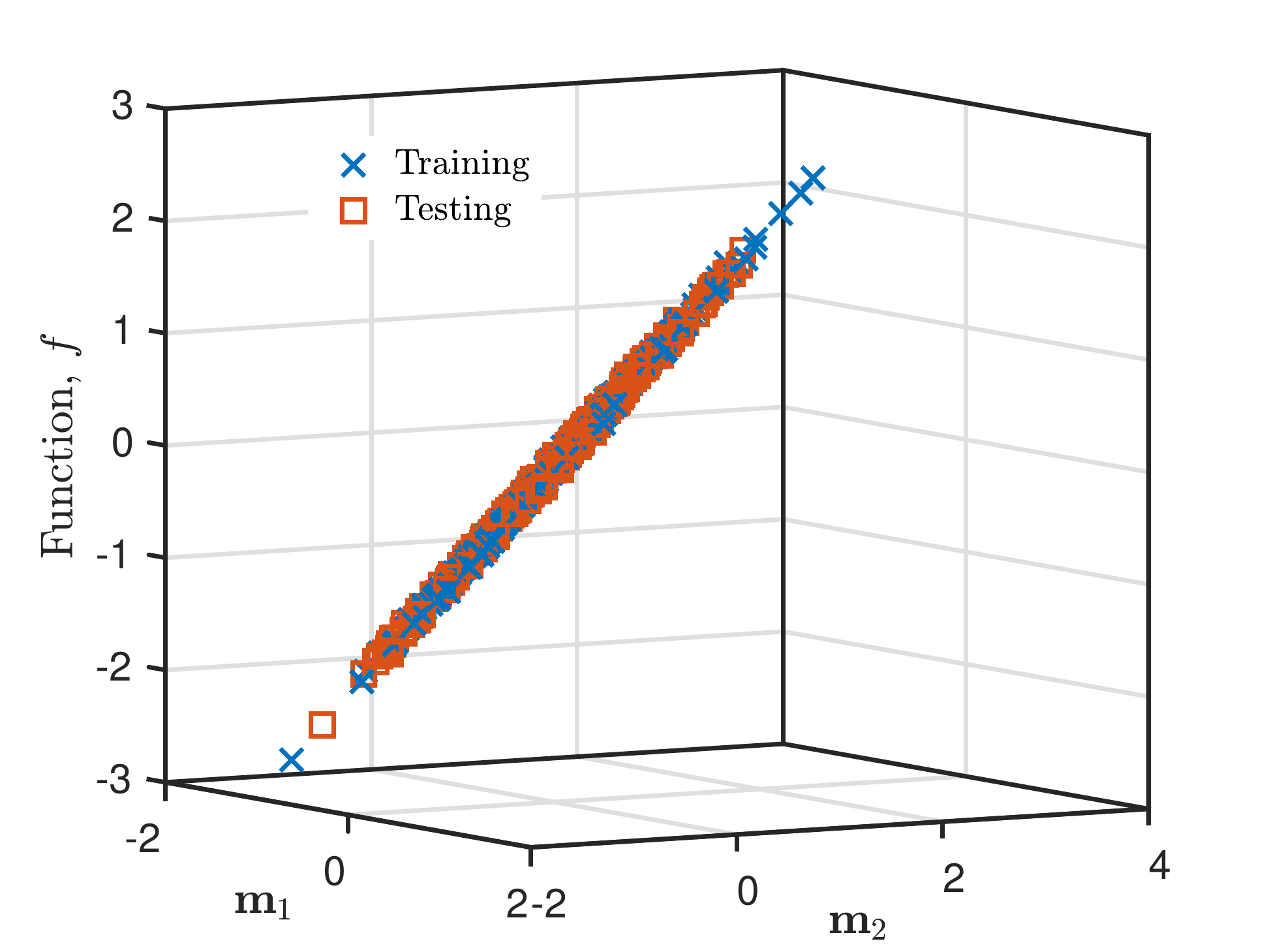}}
\subfigure[]{\includegraphics{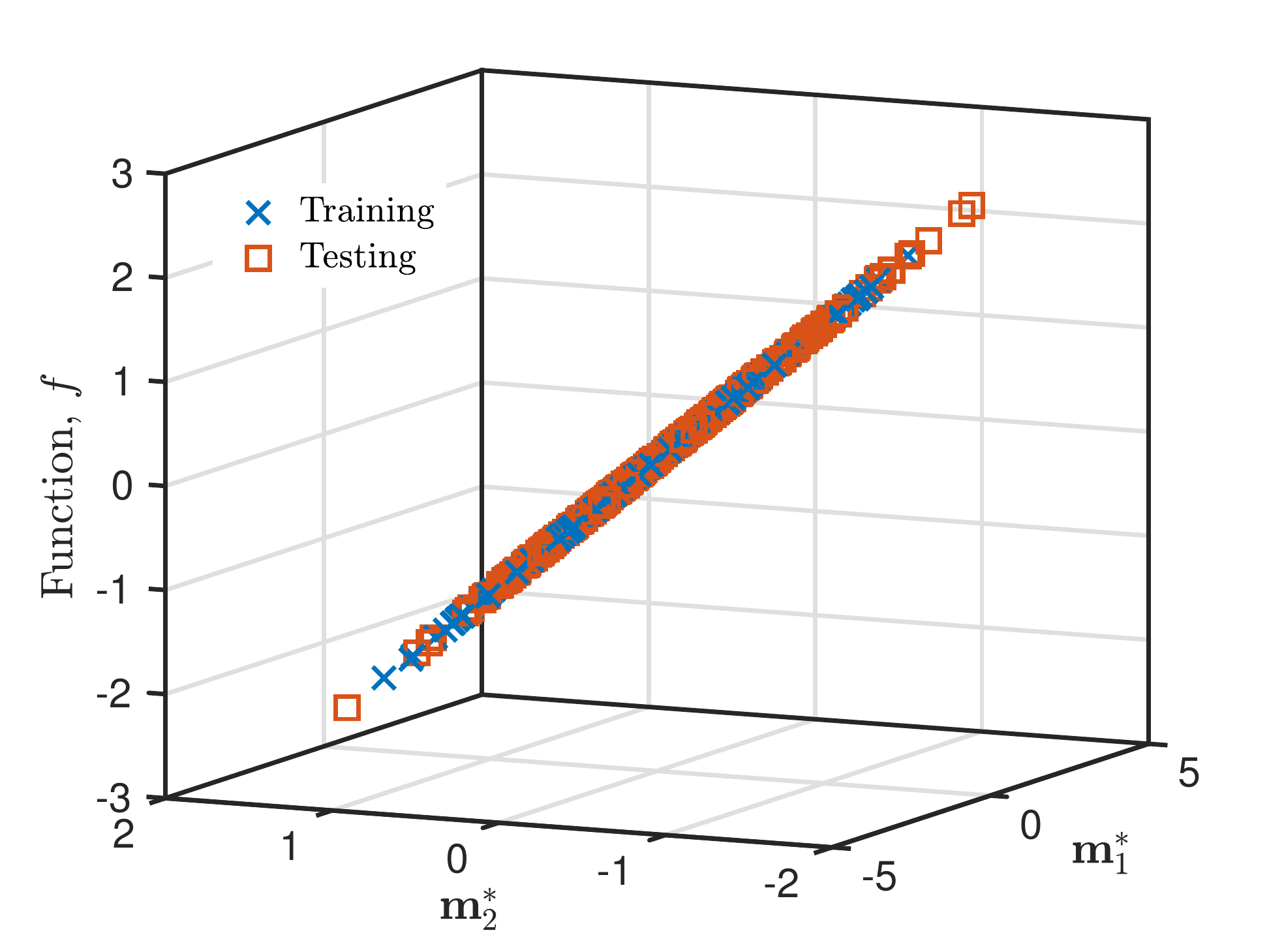}}
\subfigure[]{\includegraphics{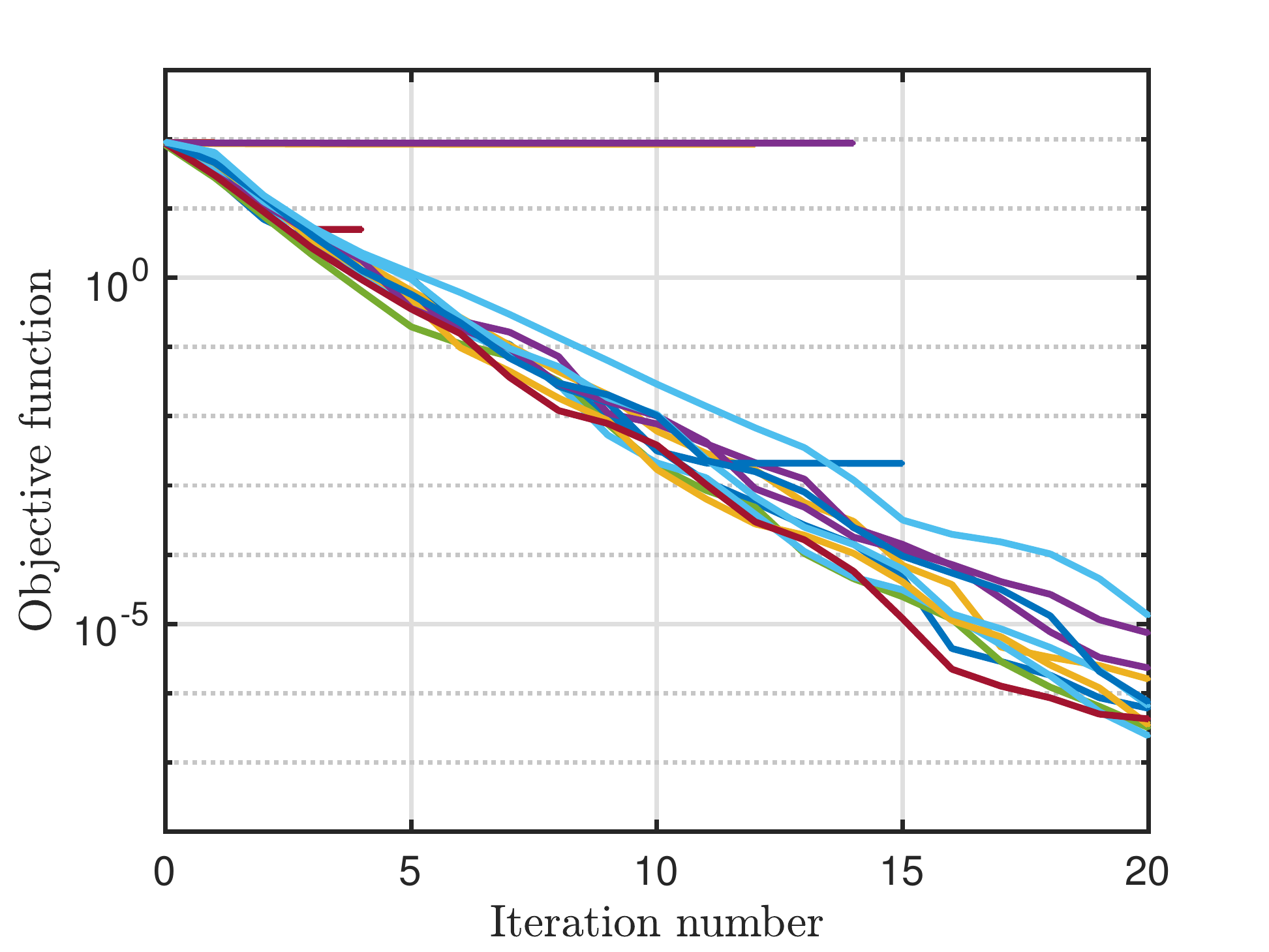}}
\subfigure[]{\includegraphics{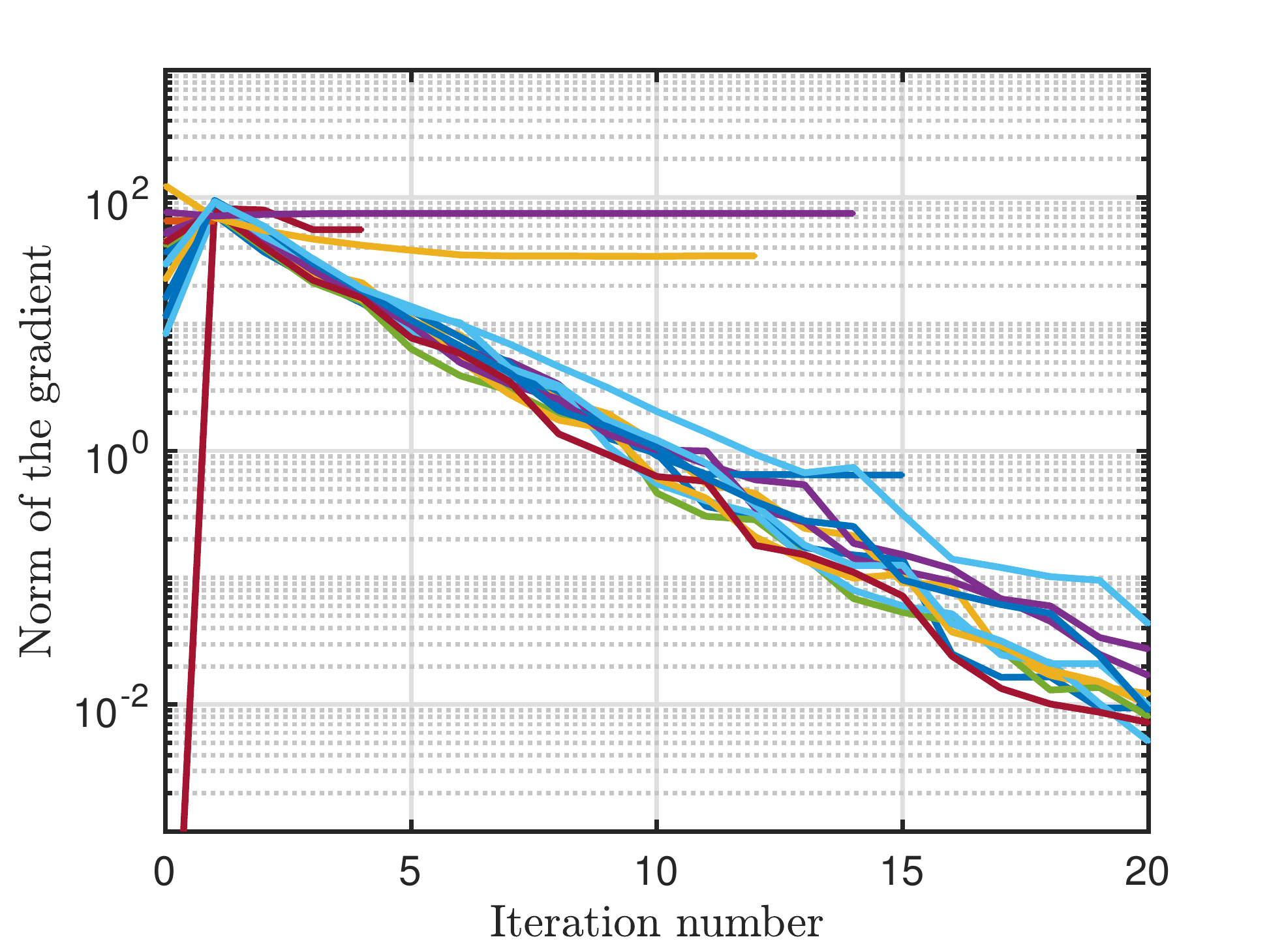}}
\subfigure[]{\includegraphics{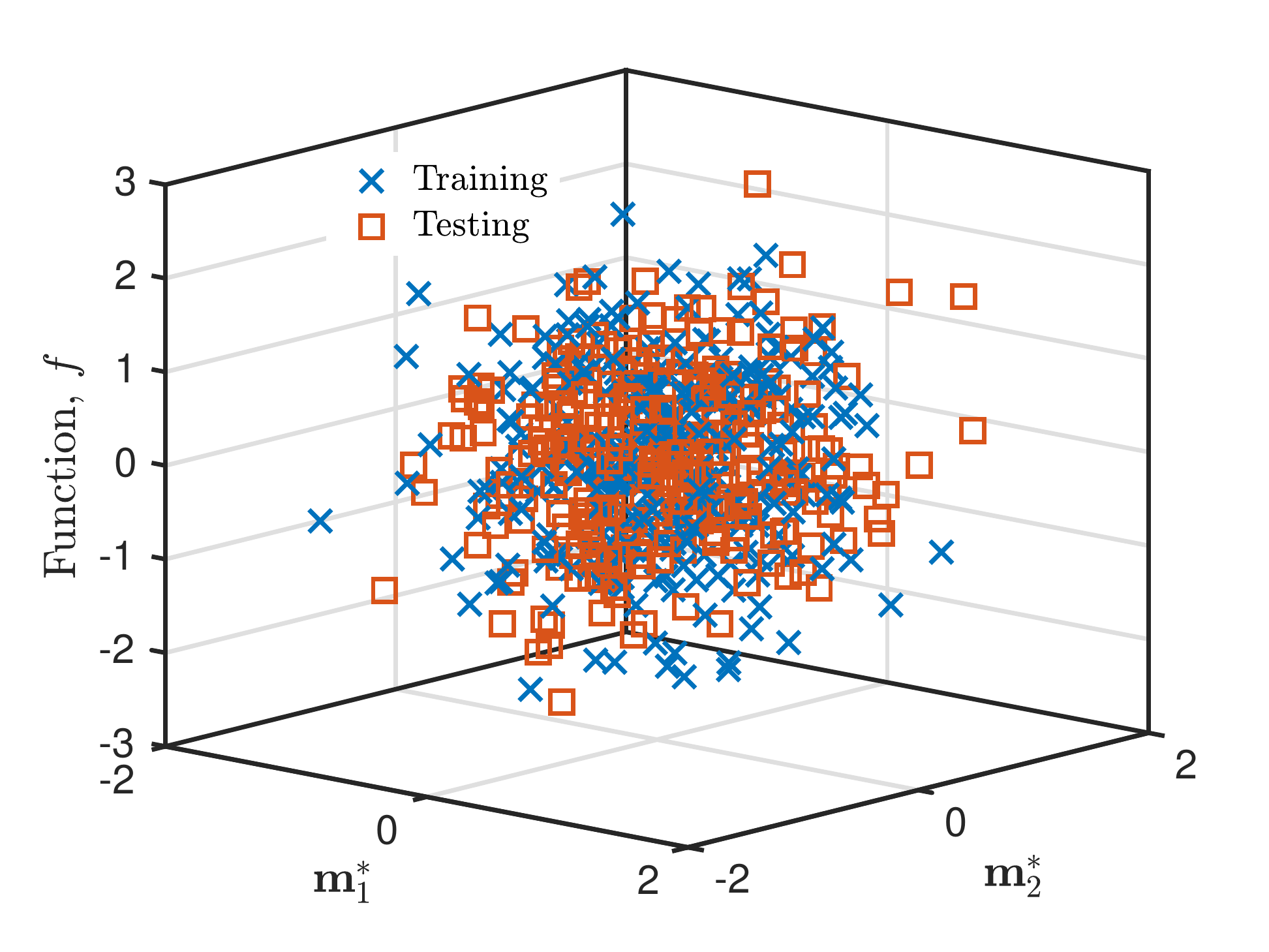}}
\end{subfigmatrix}
\caption{Results of the analytical linear ridge problem with $d=100$, $N_{test}=300$ and $N_{train}=300$ with 20 repetitions: (a) Sufficient summary plot on $\mM$; (b) sufficient summary plot on $\mM^{\ast}$ taken from the trial with the lowest value of the objective function; (c) objective function vs. iterations; (d) gradient norm vs. iterations; (e) Sufficient summary plot on $\mM^{\ast}$ taken from the trial with a high value of the objective function.}
\label{gp_100}
\end{figure}

A similar plot for the case where $d=100$ is shown in Figure~\ref{gp_compare}(b). Here we observe that on average, beyond $N_{train}=200$ we observe high recovery probabilities. Thus, we repeat the same set of experiments with $N_{test}=300$ and $N_{train}=300$ and plot the results in Figure~\ref{gp_100}. Once again our algorithm is able to identify the ridge structure associated with this problem. It should be noted that in even with 300 testing and training samples, we do encounter solutions that do not converge, as shown in Figure~\ref{gp_100}(c) and (d). Our strategy for finding a suitable solution is to repeat a single experiment---fixing the number of training and testing samples---20 times and then select the solution that yields the lowest value of the objective function. This is one of the key shortcomings of our approach, i.e., the need to run several random trials and then select a solution. Selecting one of the optimization trials with poor convergence results sufficient summary plots of the form Figure~\ref{gp_100}(e). Here it is clear that the ridge structure is not identified. 

\subsection{A bivariate normal distribution}
\begin{figure}
\begin{subfigmatrix}{2}% number of columns
\subfigure[]{\includegraphics{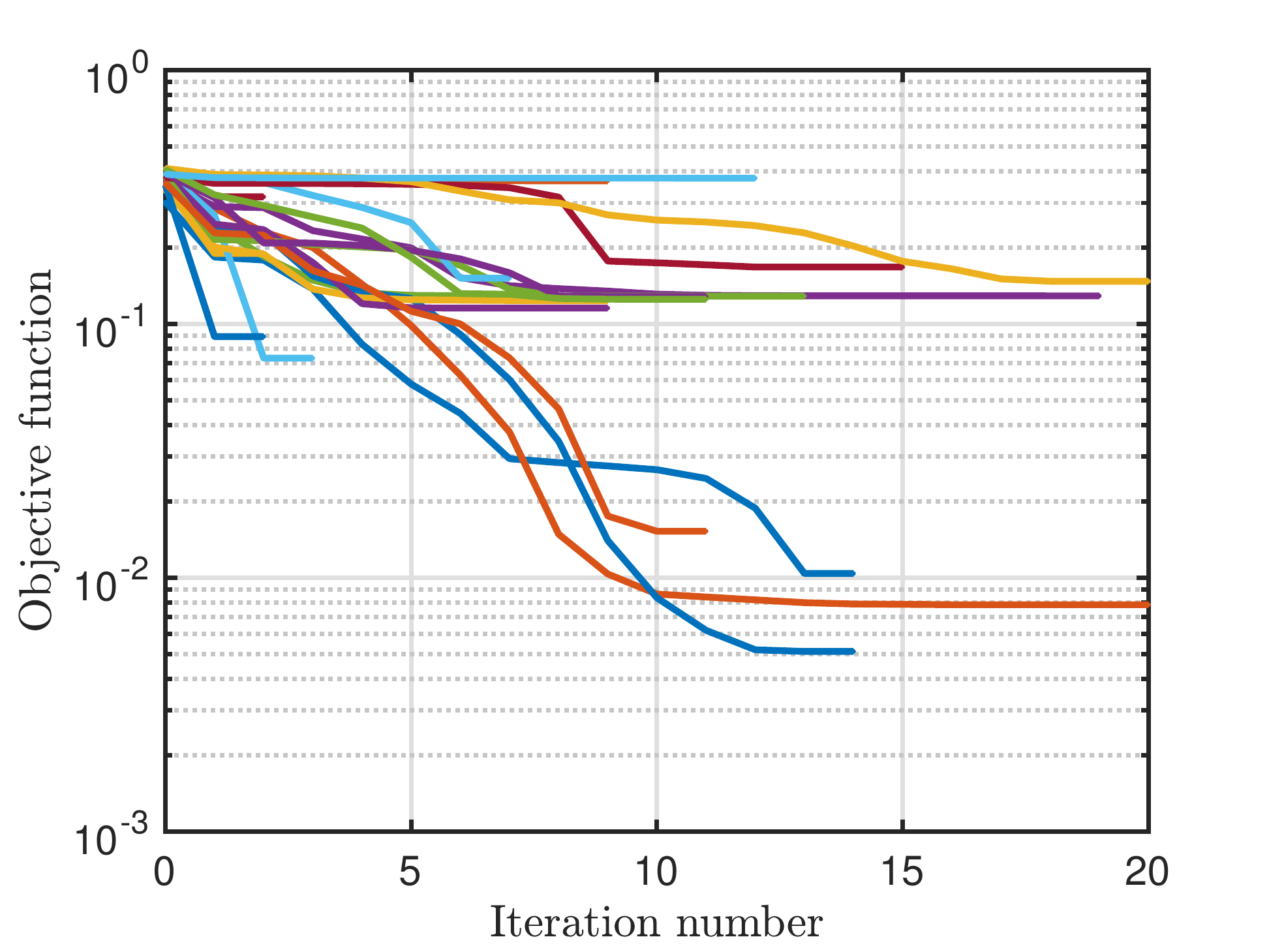}}
\subfigure[]{\includegraphics{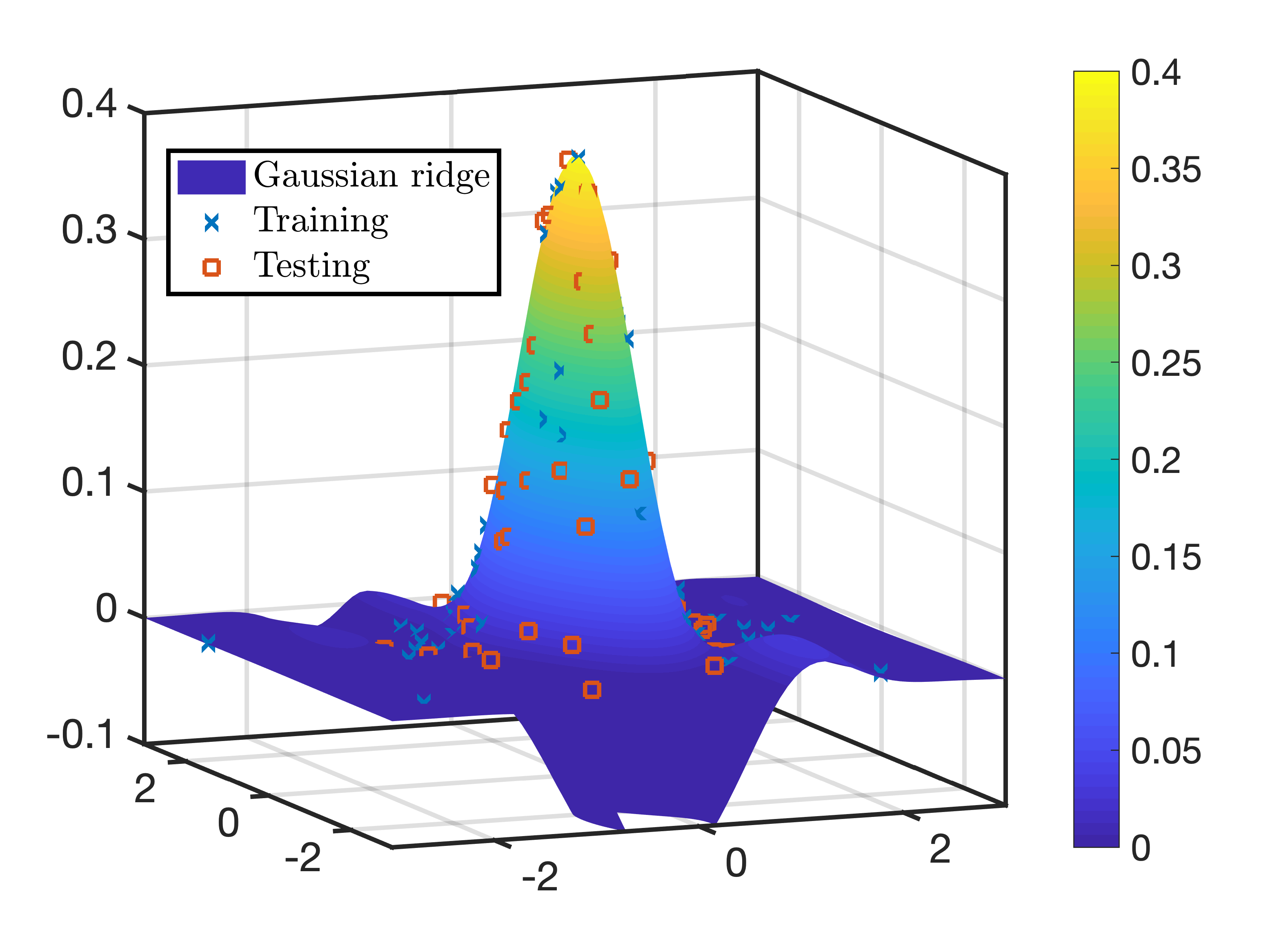}}
\subfigure[]{\includegraphics{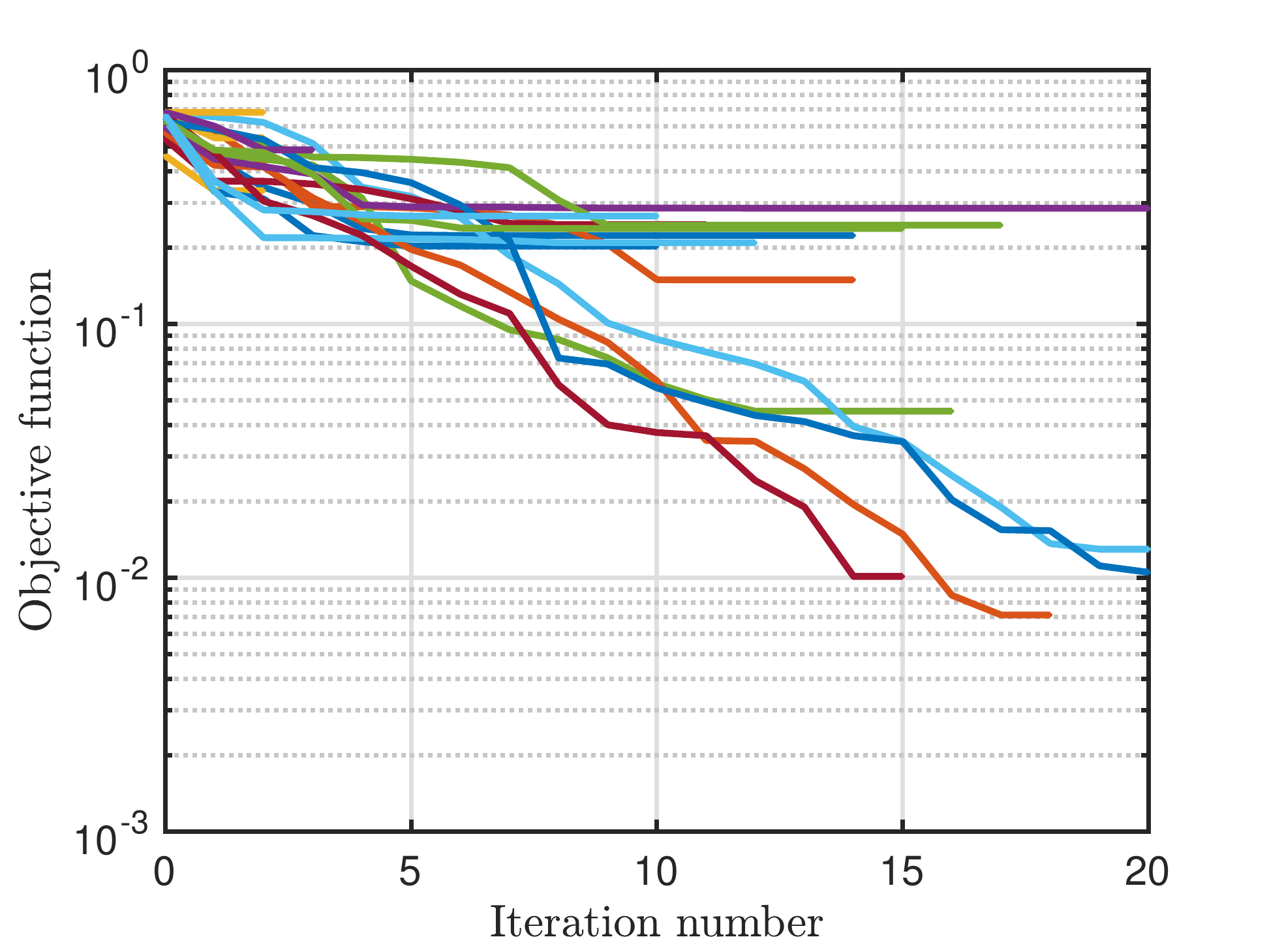}}
\subfigure[]{\includegraphics{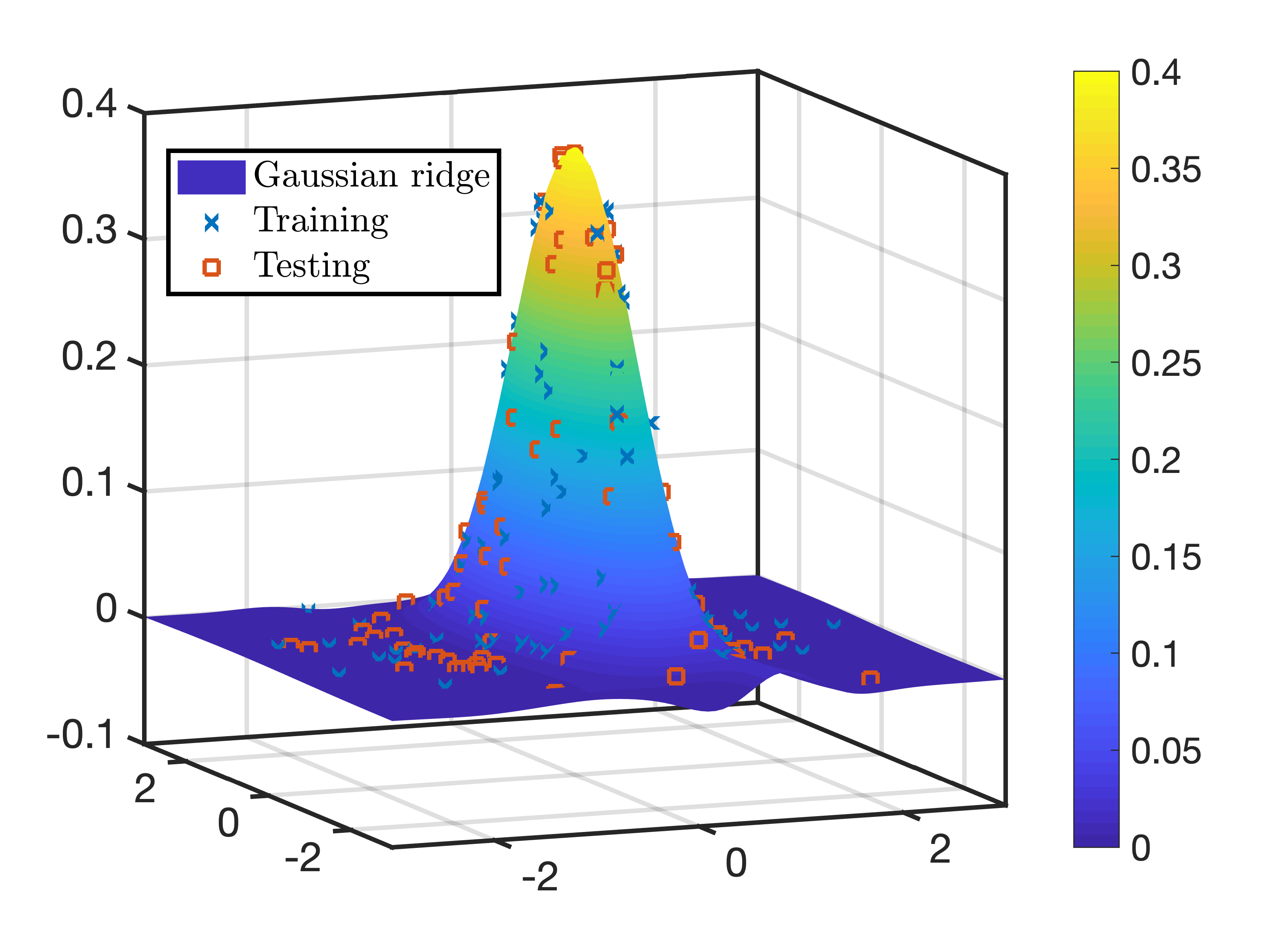}}
\subfigure[]{\includegraphics{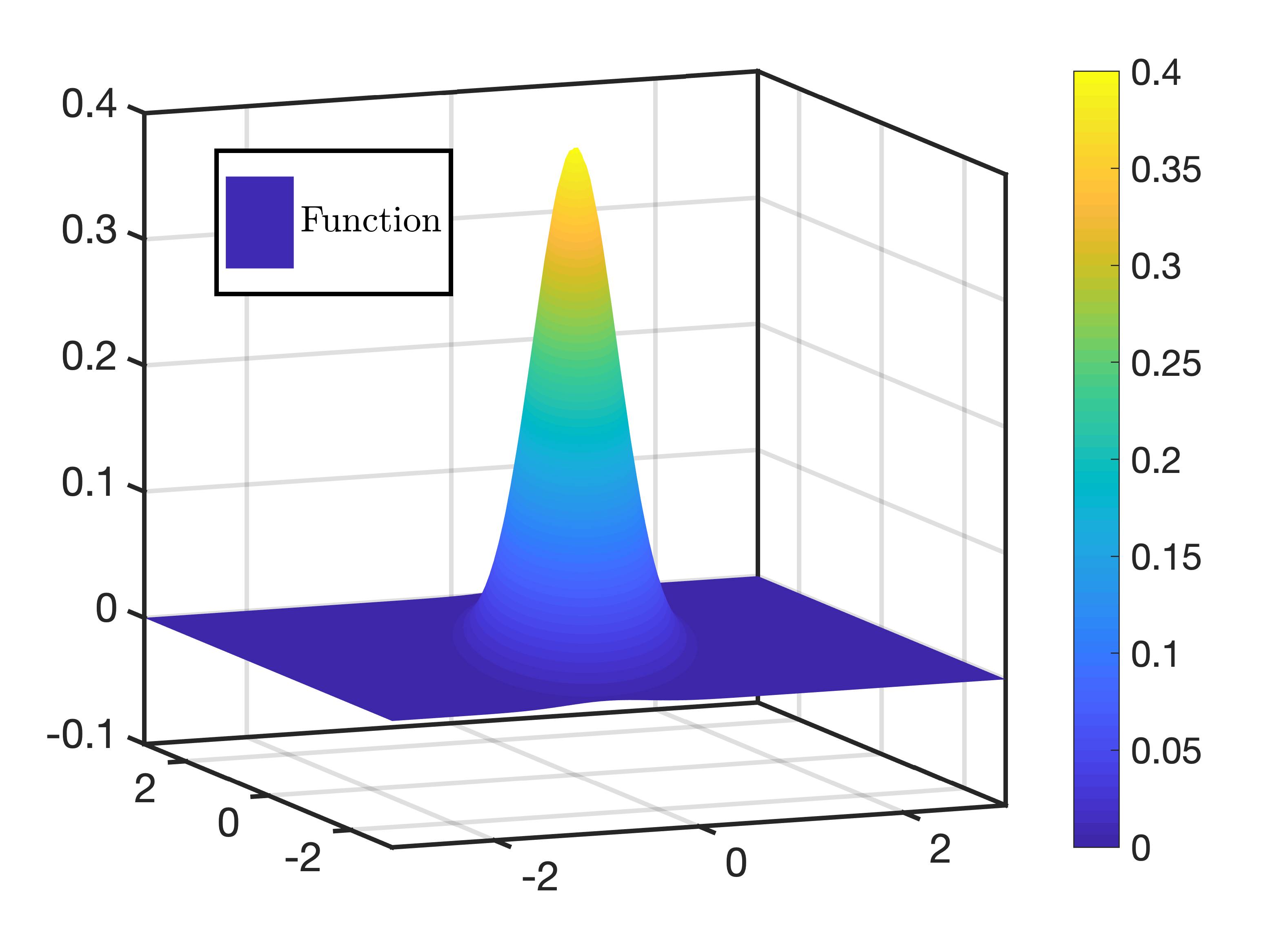}}
\end{subfigmatrix}
\caption{Results of the bivariate normal distribution problem with $d=5$ with the true function in (e) and approximations with (a, b) $N_{test}=50$, $N_{train}=50$ and (c, d) $N_{test}=100$, $N_{train}=100$.}
\label{gp_gp}
\end{figure}

In this example,  we seek to approximate the function
\begin{equation}
f\left(\mM^{T} \vx\right)=\frac{1}{\sqrt{\left| \boldsymbol{\varSigma} \right|\left(2\pi\right)^{2}}}exp\left(-\frac{1}{2} \vu \boldsymbol{\varSigma} ^{-1} \vu^{T}\right),
\end{equation}
where
\begin{equation}
\mM=\left[\begin{array}{cc}
-0.5425 & 0.0654\\
-0.6784 & -0.4931\\
-0.2381 & -0.2367\\
-0.1655 & 0.4643\\
-0.4017 & 0.6935
\end{array}\right], \; \; \; \text{and} \; \; \; \boldsymbol{\varSigma}=\left[\begin{array}{cc}
0.25 & 0.30\\
0.30 & 1.0
\end{array}\right],
\end{equation}
where identically distributed independent samples are taken from a multivariate normal distribution with zero mean and a variance of unity, i.e., $\vx \in \mathcal{N} \left(0, 1 \right)^{5}$.
Thus, our 5D function is actually a multivariate Gaussian distribution over a 2D subspace. In Figure~\ref{gp_gp} we compare the original function in (e) with two different approximations, shown in (a,b) and (c,d). These approximations use a different number of training and testing pairs, starting with 50 in (a, b) and 100 $N_{test}$ and $N_{train}$ samples each in (c,d). As before, we examine a family of 20 random trials using the same testing and training input/output pairs, and then select the solution that yields the lowest value of the objective function. Once again, it is clear that our solution strategy is able to roughly identify the ridge structure associated with the problem. 

\subsection{Turbomachinery case study}
In this subsection, we use the turbomachinery case study considered in~\cite{seshadri2017turbo} to compare the efficacy of our algorithm with a few sufficient dimension reduction techniques. We make two key claims in this subsection for this turbomachinery case study: (a) the standard deviation obtained from a Gaussian process may be used to make statements on the accuracy associated with the ridge approximation; and (b) our algorithm achieves a lower predictive variance compared to four well-known sufficient dimension reduction methods: SIR, SAVE and CR. To begin, we briefly summarize the turbomachinery case presented here.

In their work investigating the use of active subspaces for learning turbomachinery aerodynamic pedigree rules of design, the authors of~\cite{seshadri2017turbo} parameterize an aero-engine fan blade with 25 design variables---five degrees of freedom defined at five spanwise locations. Once the geometry is generated, the mesh generation and flow solver codes described in~\cite{seshadri2017turbo} are used to solve the Reynolds average Navier Stokes equations (RANS) for the particular blade design; a post-processing routine is then utilized for estimating the efficiency---our \emph{quantity of interest}. In the aforementioned paper, the authors run a design of experiment (DOE) with 548 different designs for estimating the active subspaces via a global quadratic model, which required estimating 351 polynomial coefficients. In what follows, we demonstrate the efficacy of our algorithm with far fewer samples. 

We randomly (uniformly) select $N=300$ samples from the above DOE: input/output pairs ($\vx_i \in \mathbb{R}^{25}$, $f_i \in \mathbb{R}$) for $i=1, \ldots, N$. These samples are provided as inputs to SIR, SAVE and CR for estimating a dimension-reducing subspace. The computed subspaces $\tilde{\mM} \in \mathbb{R}^{d \times 2}$ are then used for generating sufficient summary plots and for fitting a Gaussian process regression response. To ensure parity, all methods use the same 300 samples. 

\begin{figure}
\begin{subfigmatrix}{3}% number of columns
\subfigure[]{\includegraphics{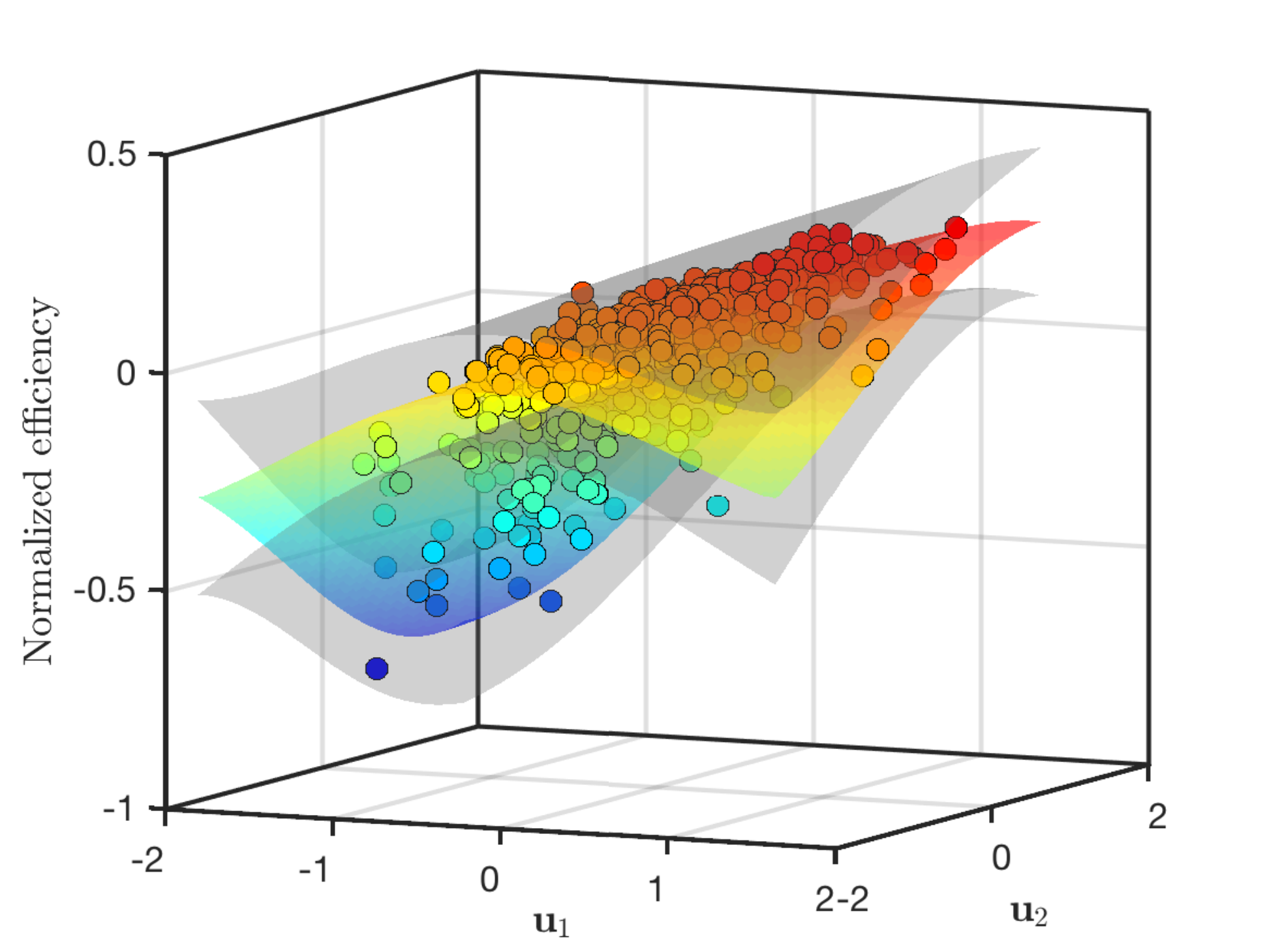}}
\subfigure[]{\includegraphics{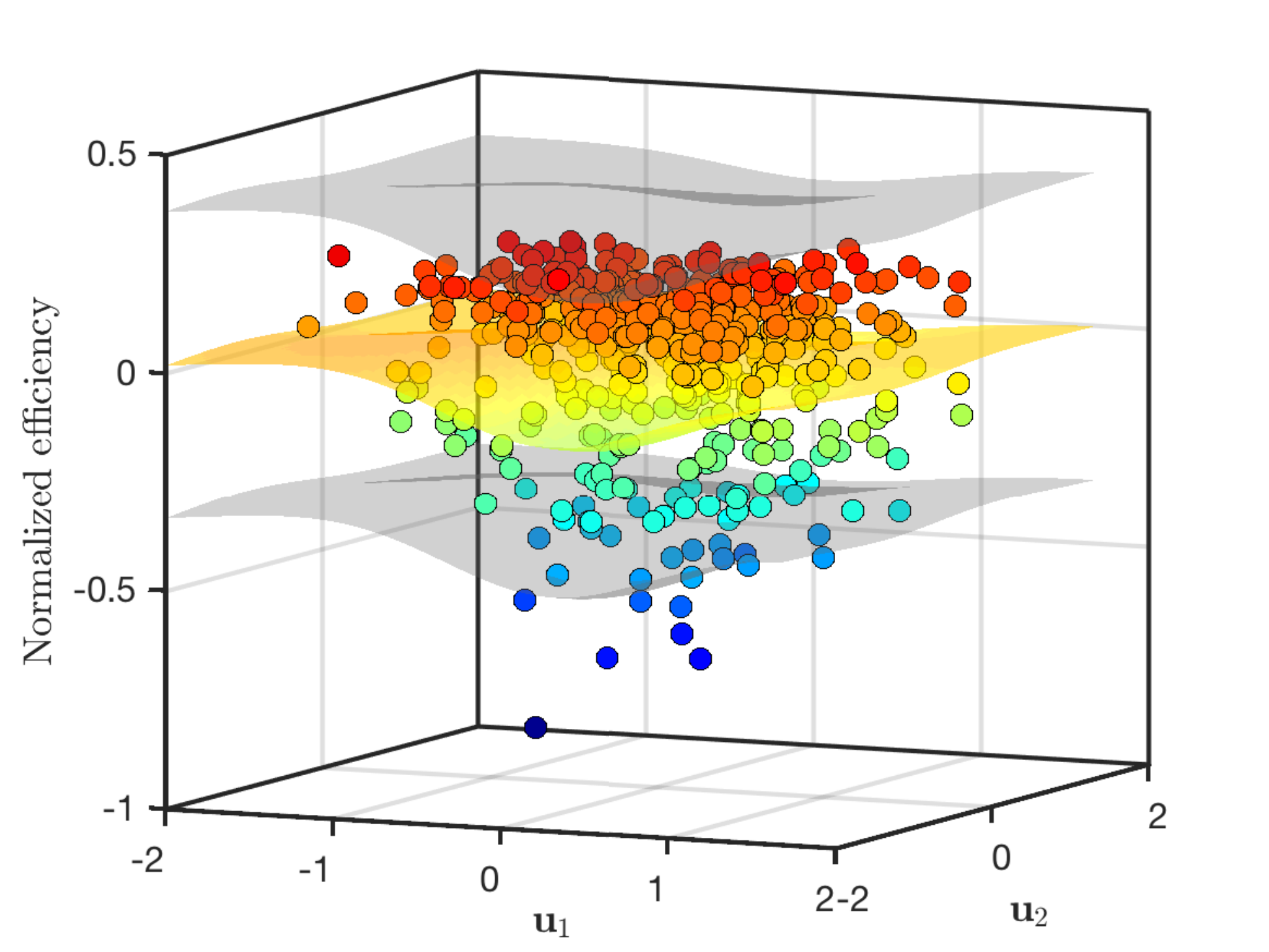}}
\subfigure[]{\includegraphics{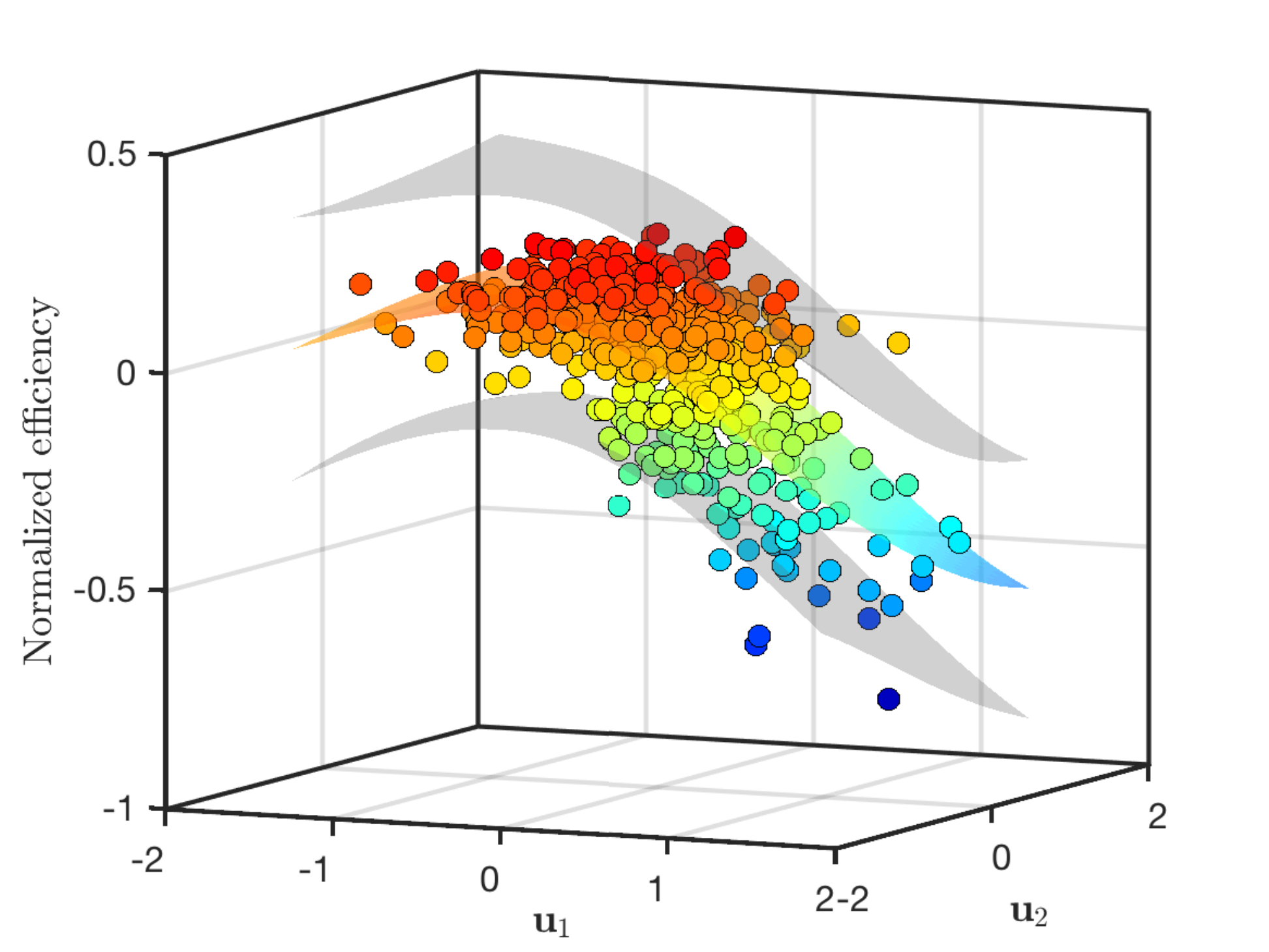}}
\subfigure[]{\includegraphics{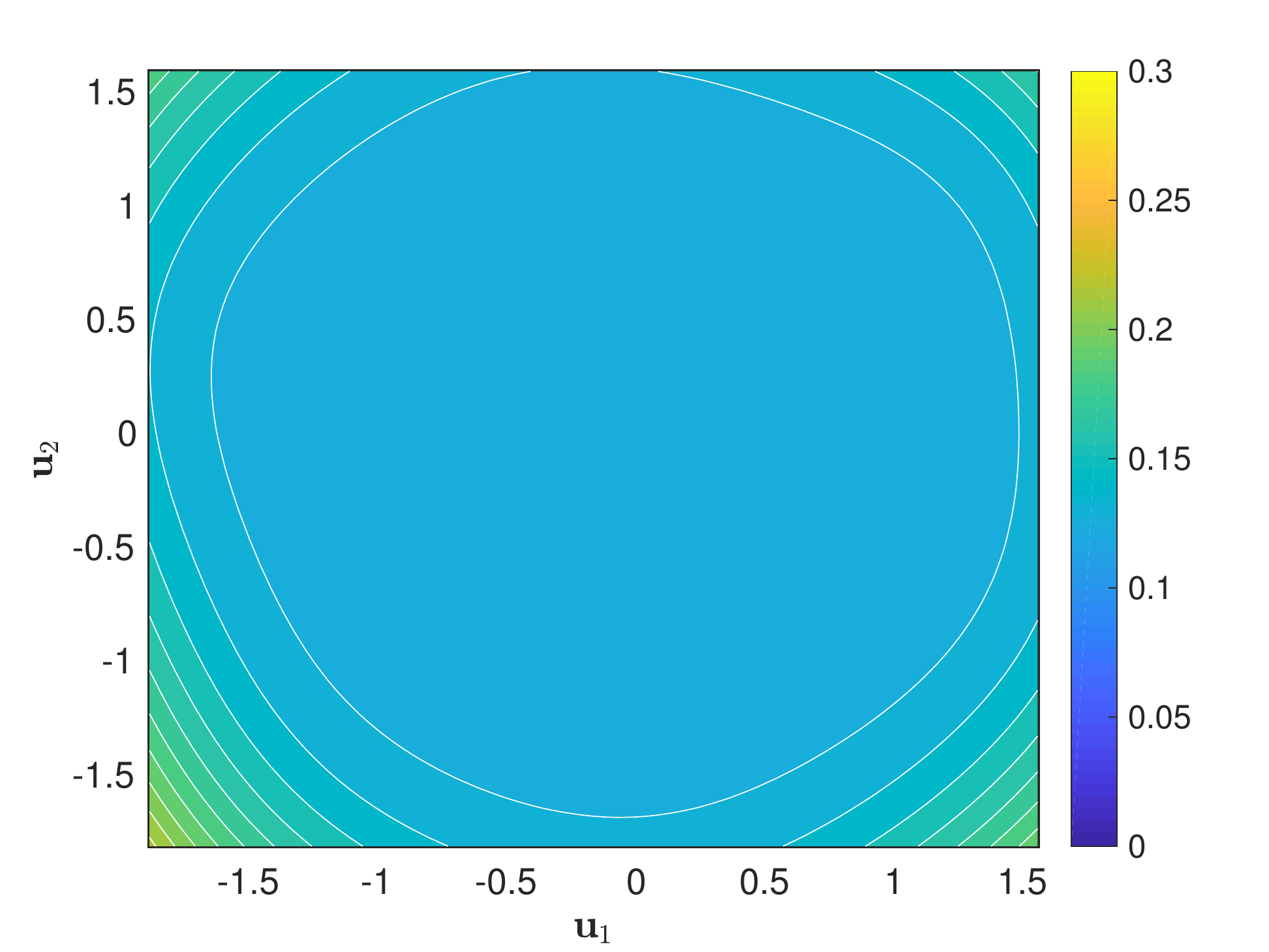}}
\subfigure[]{\includegraphics{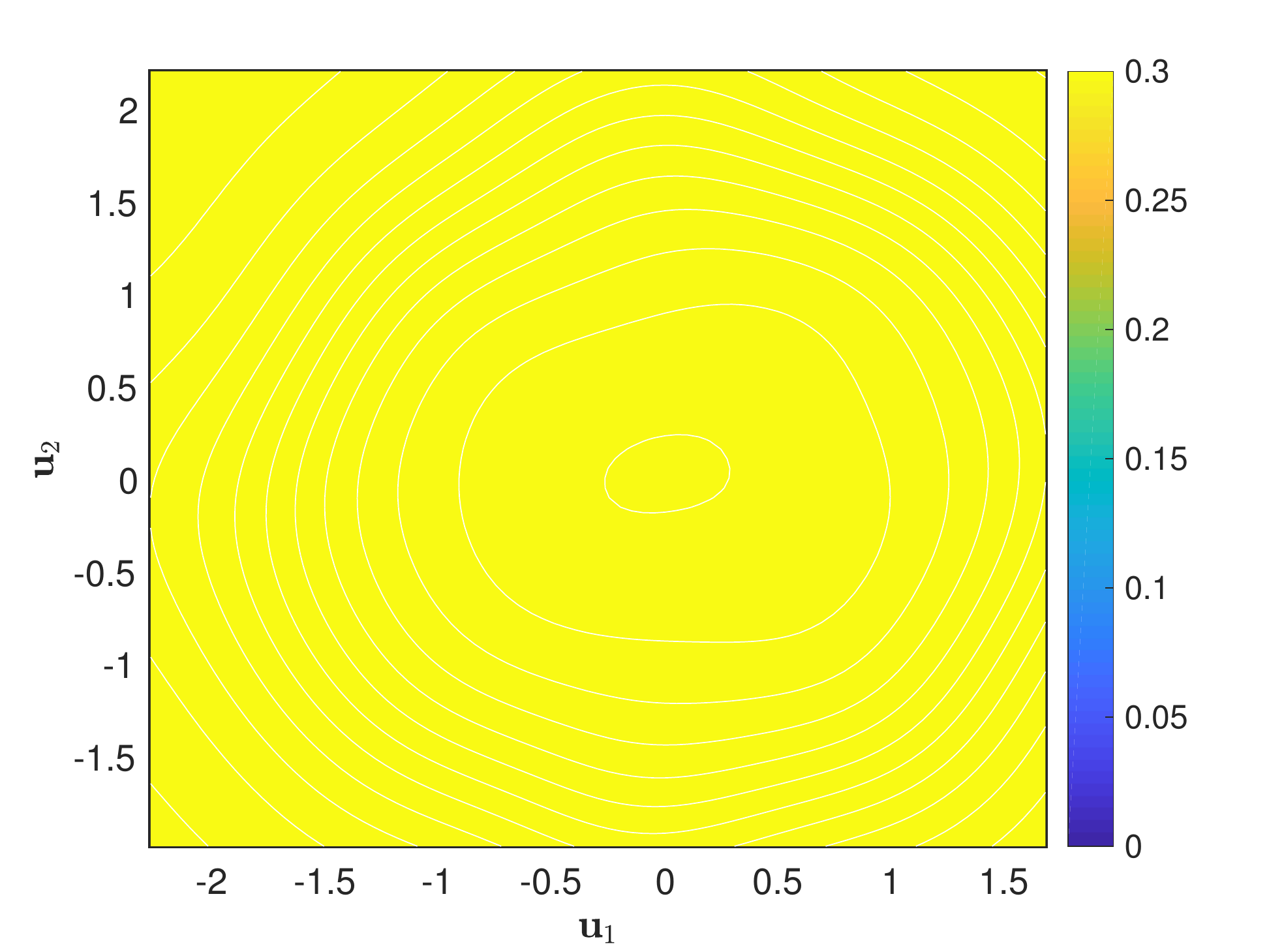}}
\subfigure[]{\includegraphics{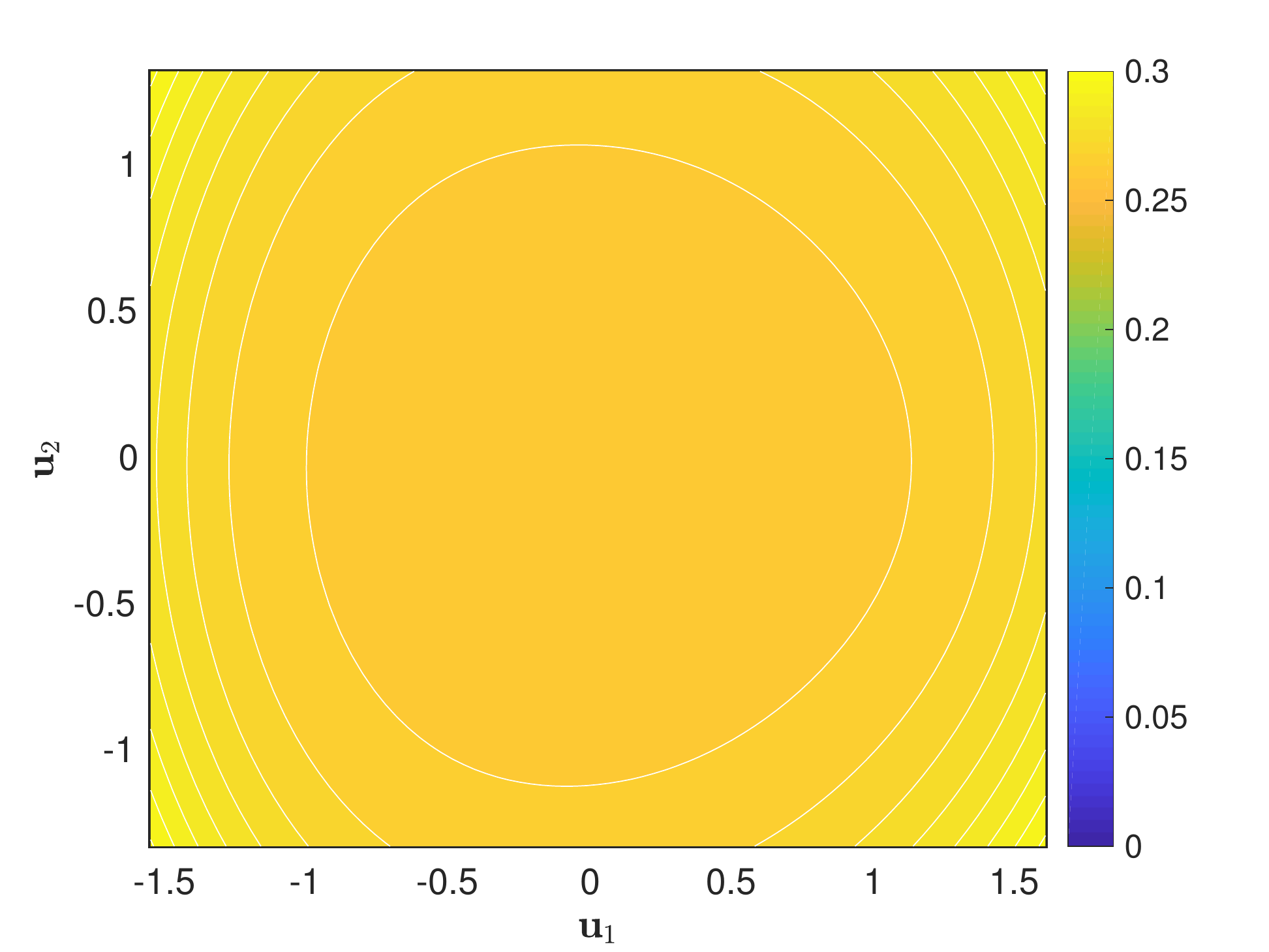}}
\subfigure[]{\includegraphics{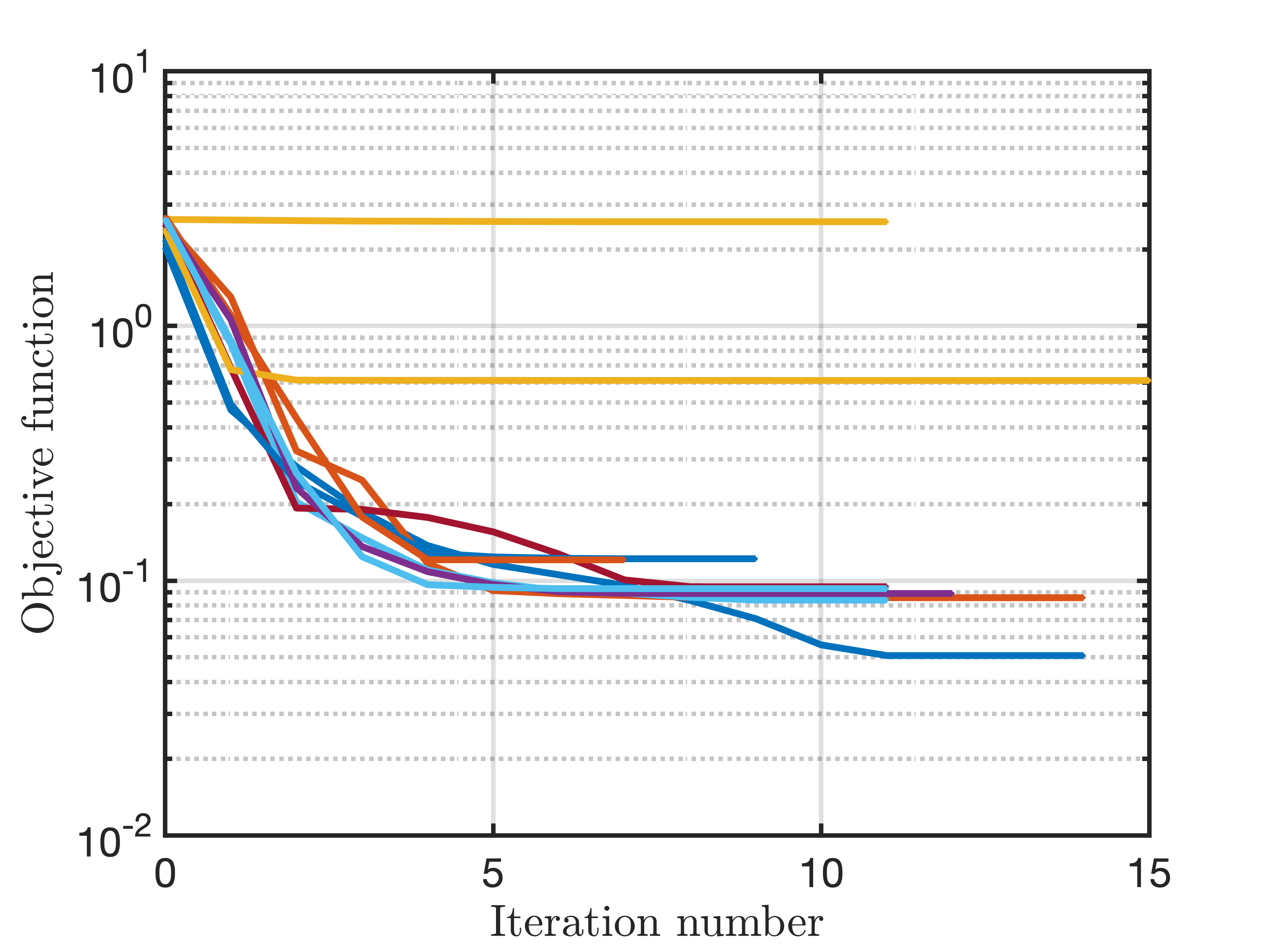}}
\subfigure[]{\includegraphics{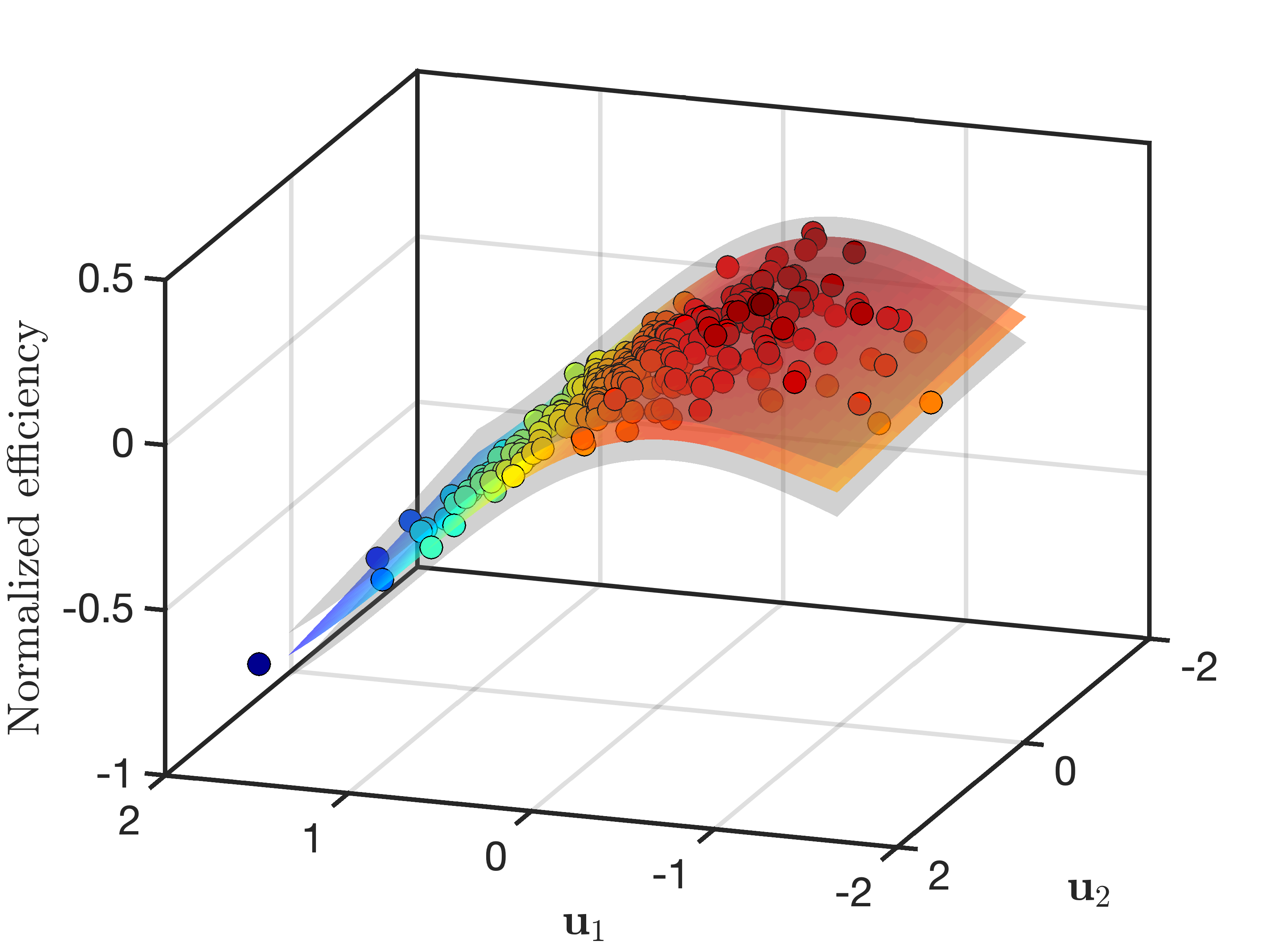}}
\subfigure[]{\includegraphics{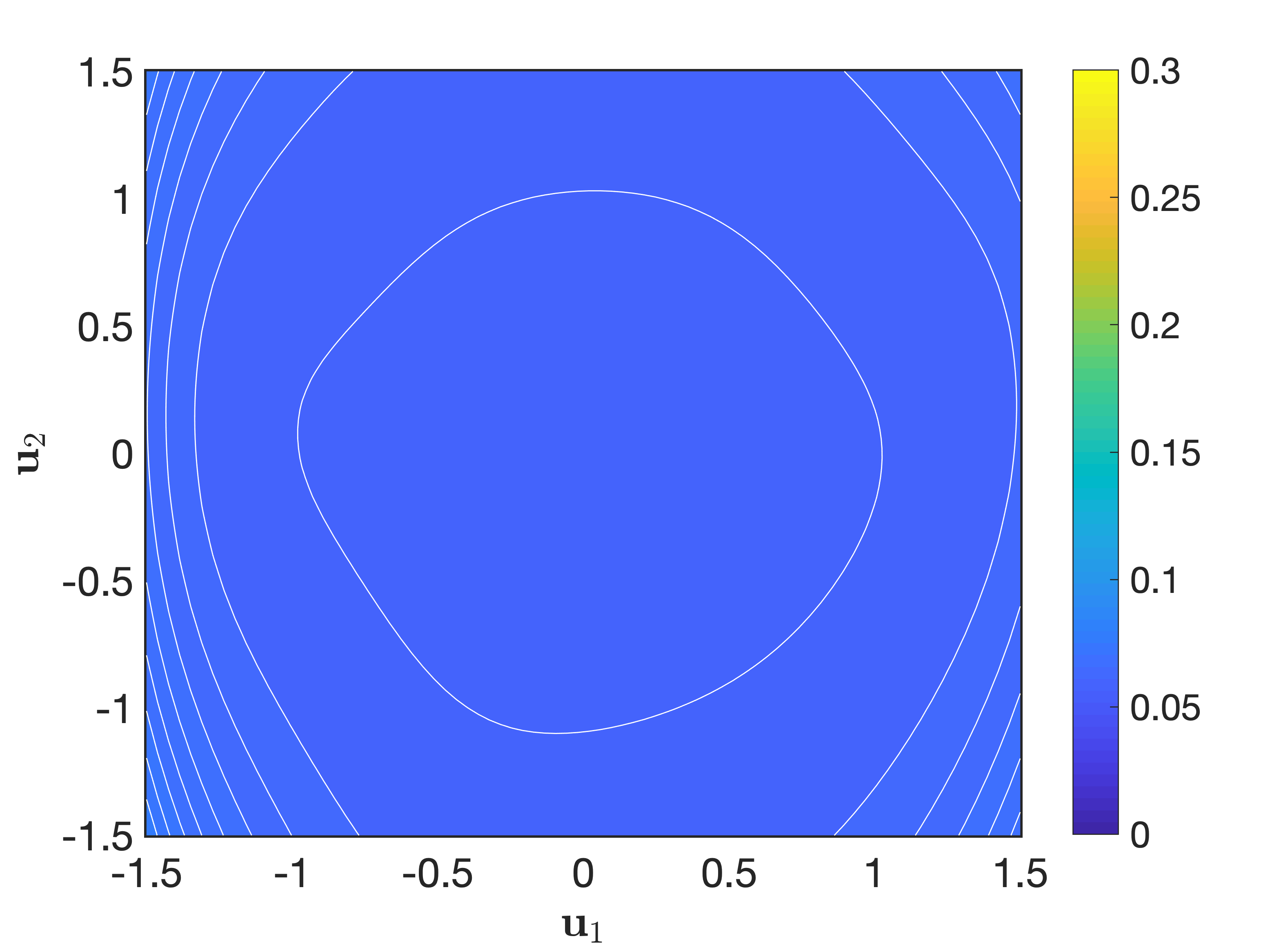}}
\end{subfigmatrix}
\caption{Turbomachinery case study results for estimating the ridge subspace followed by Gaussian process regression for estimating the mean and 2$\sigma$ response surfaces. Sufficient summary plots for SIR, SAVE and CR are shown in (a, b and c); corresponding $2\sigma$ contours are shown in (d, e and f). Optimization convergence for 20 random trails for our algorithm is shown in (g), sufficient summary plot in (h) and the corresponding $2\sigma$ plot in (i).}
\label{gp_sir_save}
\end{figure}

To ascertain how appropriate the choice of a particular $\tilde{\mM}$ is, we use all 548 DOE points in the sufficient summary plots and in the Gaussian process regression (even though the subspaces $\tilde{\mM}$ are estimated with strictly 300 samples). Figures~\ref{gp_sir_save} (a, b amd c) are the computed sufficient summary plots for SIR, SAVE and CR. Here the vertical axis represents the normalized efficiency, while the horizontal axis for a particular design $\vx_i$ is given by
\begin{equation}
\vu_{1} = \left( \tilde{\mM}(:, 1) \right)^T \vx \; \; \; \text{and} \; \; \; \vu_{2} = \left( \tilde{\mM}(:, 2) \right)^T \vx,
\end{equation}
where the notation $(:,k)$ denotes the $k$-th column of $\tilde{\mM}$. Figures~\ref{gp_sir_save} (d, e and f) show contours of the $2\sigma$ values obtained from the Gaussian process response surface. It is clear from these plots that SIR yields lower $2\sigma$ values on average. SAVE and CR fail to identify any low-dimensional structure. In contrast, the result of our algorithm is shown in Figure~\ref{gp_sir_save}(g-i). The $2\sigma$ contours shown in Figure~\ref{gp_sir_save}(i) are by far the lowest, and one can clearly see that all the data lies on a quadratic manifold. This result was obtained by selecting the subspace that yielded the lowest value of the objective function after 20 trials (see Figure~\ref{gp_sir_save}(g)).

\section{Conclusion}
In this paper we introduce a new algorithm for ridge function approximation tailored for subspace-based dimension reduction. Given point evaluations of a model, our algorithm approximates the ridge subspace $\mM$ by minimizing a quadratic form of the function and its best ridge approximation. Throughout this paper, we assume our ridge function is the posterior mean of a Gaussian process. Our results for the examples considered demonstrate that: (i) our algorithm is able to obtain near local minima recovery; and (ii) our algorithm offers better results compared with the four sufficient dimension reduction techniques considered in this paper. 

\section{Acknowledgements}
The authors are grateful to Paul Constantine for insightful discussions. The first author would like to acknowledge the support of a Rolls-Royce fellowship. The second author would like to acknowledge the financial support of Magdalene College, Cambridge. 

\bibliography{references}
\bibliographystyle{siamplain}

\end{document}